\documentclass[final]{IEEEtran}
\ifCLASSINFOpdf
\else
\fi

\usepackage{amsthm,amssymb,graphicx,multirow,amsmath,color,amsfonts}
\usepackage[update,prepend]{epstopdf}
\usepackage[noadjust]{cite}
\usepackage{caption}
\usepackage{tabulary}
\usepackage{bbm} 
\usepackage{multirow}
\usepackage{comment}
\usepackage{mathtools}
\usepackage{tabularx}
\usepackage[caption=false,font=normalsize,labelfont=sf,textfont=sf]{subfig}
\usepackage{balance}
\usepackage{enumitem,kantlipsum}

\usepackage[dvipsnames]{xcolor}
\usepackage[switch]{lineno}

\newcolumntype{s}{>{\hsize=.8\hsize}X}
\newcolumntype{b}{>{\hsize=1.2\hsize}X}

\DeclarePairedDelimiter\abs{\lvert}{\rvert}%
\DeclarePairedDelimiter\norm{\lVert}{\rVert}%

\makeatletter
\let\oldabs\abs
\def\abs{\@ifstar{\oldabs}{\oldabs*}}
\let\oldnorm\norm
\def\norm{\@ifstar{\oldnorm}{\oldnorm*}}
\makeatother

\def\nb0{{\mathbf{0}}}
\def\nb1{{\mathbf{1}}}

\newtheorem{definition}{Definition}

\newtheorem{prop}{Proposition}

\allowdisplaybreaks 

\usepackage{setspace}	

\begin{document}
\pagenumbering{gobble}
\title{Toward Sustainable Transportation: Accelerating Vehicle Electrification with Dynamic Charging Deployment
}
\author{
Duc Minh Nguyen, \textit{Student Member, IEEE}, Mustafa A. Kishk, \textit{Member, IEEE}, and Mohamed-Slim Alouini, \textit{Fellow, IEEE}
\thanks{
Copyright \copyright{} 2022 IEEE. Personal use of this material is permitted. However, permission to use this material for any other purposes must be obtained from the IEEE by sending a request to pubs-permissions@ieee.org.

Duc Minh Nguyen and Mohamed-Slim Alouini are with Computer, Electrical and Mathematical Science and Engineering Division, King Abdullah University of Science and Technology (KAUST), Thuwal  23955-6900, Saudi Arabia (email: \{ducminh.nguyen; slim.alouini\}@kaust.edu.sa).

Mustafa A. Kishk was with KAUST until January 2022. Currently, he is with the Department of Electronic Engineering, National University of Ireland, Maynooth, W23 F2H6, Ireland. (email: mustafa.kishk@mu.ie).} 
}

\maketitle
\begin{abstract}
Electric vehicles (EVs) are being actively adopted as a solution to sustainable transportation. However, a bottleneck remains with charging, where two of the main problems are the long charging time and the range anxiety of EV drivers. In this research, we investigate the deployment of dynamic charging systems, i.e., electrified roads that wirelessly charge EVs on the go, with a view to accelerating EV adoption rate. We propose a traffic-based deployment strategy, statistically quantify its impact, and apply the strategy to two case studies of real traffic in New York City (USA) and Xi'an (China). We find that our analytical estimates not only closely match the real data, but they also suggest that dynamic charging considerably extends the driving range of popular EV models in urban mobility. For example, when only 5\% of the existing roads in New York City are equipped with this technology, an EV model such as the Nissan Leaf will approximately maintain its battery level without stopping to recharge. If the percentage of charging roads is increased to 10\%, then the Leaf will gain nearly 10\% of its battery after every 40 kilometers of driving. Our framework provides a solution to public and private organizations that support and facilitate vehicle electrification through charging infrastructure.
\end{abstract}
\IEEEpeerreviewmaketitle
\section{Introduction}
\label{introduction}
In recent years, an increasing proportion of the general public gets to know and supports the idea of sustainable development, which is about meeting the needs of the present without compromising the ability of future generations to meet their needs. Therefore, the global trend toward sustainability has spread to several aspects of modern lives, e.g., production, consumption, and transportation. In fact, transportation contributes a significant portion to the global greenhouse-gas emission. Current transportation systems are accounted for 20\%-25\% of the world's energy consumption and carbon dioxide emissions~\cite{internationalenergyoutlook,trackingTransport}. In 2018, the carbon dioxide emission from transportation in the United States (US) was more than from electricity, industry, agriculture, and any other sector~\cite{inventoryofUSgreenhouse}. Furthermore, the primary cause of greenhouse-gas emission from transportation is light-duty vehicles~\cite{inventoryofUSgreenhouse}. Thus, to reduce the amount of greenhouse-gas emission, we need a solution to the current diesel and gasoline personal vehicles.

A solution that is gaining popularity is electric vehicles (EVs). EVs not only reduce our reliance on fossil fuels, cut down on polluting emissions, but also improve public health and foster economic growth~\cite{netemissionreduction,transportationemissions}. The global EVs sales topped 2.1 million USD in 2019, with China, the European Union, and the US being the three largest markets~\cite{globalevoutlook}. Even though the EV market share is growing fast, reaching 2.6\% in 2019, it is still much smaller than that of internal combustion engine vehicles (ICEVs). Two of the main reasons that prevent EVs from becoming the general public's choice are the long charging time and the range anxiety problem. In particular, most current EV models rely on batteries that get charged at charging stations. However, the charging time of EVs at those charging sites is still way more than the refueling time of ICEVs at gas stations, even with the latest fast-charging technology. In addition, EVs may even run out of power before reaching one of the charging stations. To compensate for those two problems, current EV manufacturers focus on producing large-capacity batteries with fast-charging technology. Nevertheless, those are the exact components leading to the high price of EVs, making them less affordable to the general public~\cite{electricvehiclesbatteries}. Moreover, producing large-capacity batteries in large quantities and replacing those batteries within the lifetime of EVs put significant strain on the global lithium supply, leading to possible shortage~\cite{impactoftransport}. 

\begin{figure}[t]
\centering
\includegraphics[width=1\columnwidth]{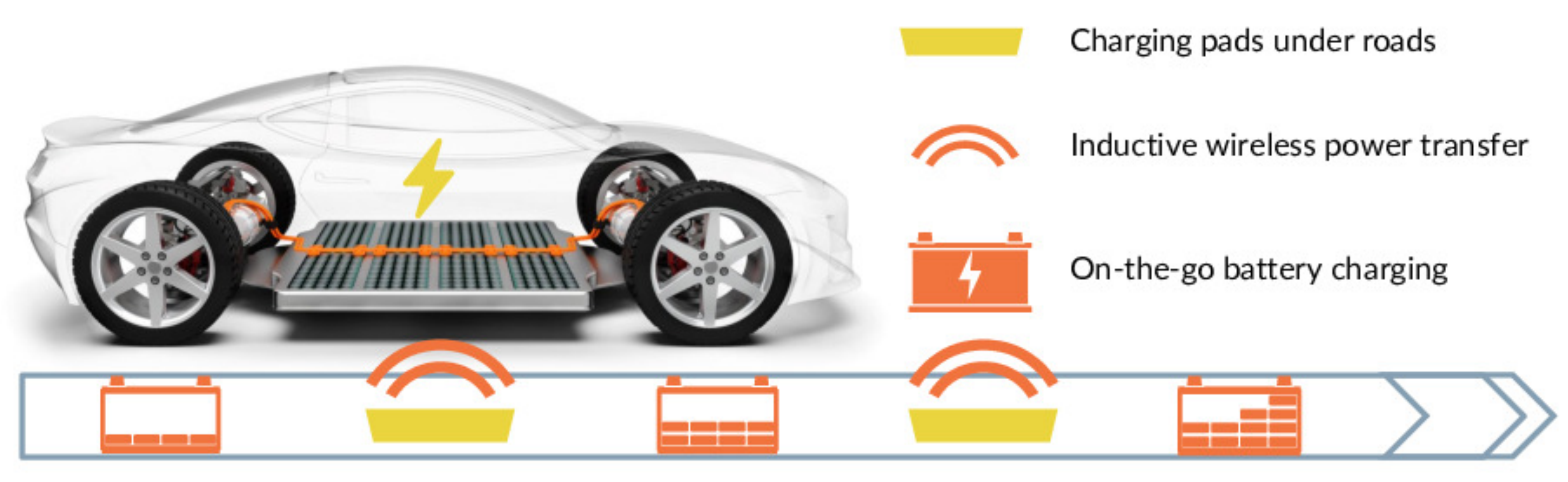}
\caption{An illustration of the dynamic charging system.}
\label{dynamic}
\end{figure}

An emerging technology that mitigates the problems with charging stations and EV batteries is dynamic charging, i.e., electrified roads that charge vehicles as they drive~\cite{reviewofstaticanddynamic}, as illustrated in Fig.~\ref{dynamic}. This system consists of charging pads installed under the roads and receivers installed at the bottom of the EVs. The charging pads use electricity to create an alternating electromagnetic field with a receiving coil. The receiving coil of an EV converts the electromagnetic field into electricity that can charge a battery or power a motor. The technology behind dynamic charging has been thoroughly studied at various institutes in the world~\cite{advancesinwirelesspowertransfer,theinductivepowertransferstory,oakridgenationallab}, and its samples have been demonstrated by many companies including industry leaders~\cite{witricity,scania,renault,electreon}. Dynamic charging gives EV drivers the option to visit charging stations less often and increases their driving range. More importantly, it reduces the pressure to produce EV models with large-capacity batteries and fast charging. Instead, EVs manufacturers may opt for smaller-size batteries with more frequent short-charging cycles, which have been shown to possess several benefits. First, they reduce the retail prices for EVs. Second, in~\cite{chargingautomationforev}, it is found that operating the batteries with shallow and frequent charges/discharges can extend the lifetime of the batteries by up to 2.9 times. Due to this enormous saving in battery cost, which surpasses the power track installation cost, over the lifetime of the batteries, dynamic charging is much more economical than charging stations~\cite{predictivemodeling,economicanalysisofdynamiccharging}. Last but not least, it is presented in~\cite{predictivemodeling} that a 30\% battery degradation leads to an 11.5–16.2\% increase in energy consumption and greenhouse-gas emissions per km. Thus, by extending the lifetime of the batteries, dynamic charging keeps EVs a green means of transportation for a more extended period.


Given the great potential of dynamic charging to accelerate the EV adoption rate and the fact that several modern cities have planned to ban diesel and gasoline cars by 2030~\cite{deloitte}, we are motivated to study the problem of deploying dynamic charging in metropolitan cities. In more detail, we seek the best plan to deploy dynamic charging systems, i.e., charging roads, measure the deployment impact into quantitative metrics, and verify our analytical results with real case studies in urban cities. 

Before elaborating on our framework, we emphasize that our solution has a big market of urban planners, city policy-makers, infrastructure construction firms, and EV manufacturers that seek insights about dynamic charging deployment. This emerging market is expected to grow quickly over the next decades because of the following reasons. First, the market for EVs is expanding fast and is forecast to reach 234 million USD by 2027~\cite{wirelesschargingmarket}. Second, several countries have announced their strong commitment to sustainability. For example, both the EU and the US aim to reduce carbon dioxide emissions by 50\% by 2030~\cite{neweutarget, factsheet}. Third, pilot programs about dynamic charging are being actively run in various test sites in the world. For instance, Renault is examining the technology with 32 partners in Europe and expects to fully incorporate the dynamic charging capability in its vehicles by 2030~\cite{electreon}. Last but not least, the EV transformation has received generous funding from government officials and private corporations, e.g., the US administration, Ford, and Hyundai all plan to heavily invest in EV infrastructure and production~\cite{biden,ford,hyundai}. 

In summary, with a view to facilitating the adoption of EVs worldwide, we present a plan to deploy and quantify the impact of a new kind of charging facility for EVs, i.e., dynamic charging systems, in metropolitan cities. Our main contributions are summarized as follows:
\begin{itemize}
    \item We investigate the spatial distribution of commuting trips and propose that the charging roads should be deployed accordingly to maximize road usage and utilization.
    \item We quantify the impact of deployment through two statistical metrics: the distribution of the distance to the nearest charging road and the distribution of trip portion traveled on the charging roads. 
    \item We demonstrate our charging road deployment strategy on case studies of New York City (NYC), USA, and Xi'an, China, in which we combine real traffic data with actual road network data to confirm our analytical results and deduce important implications on the changes of EV battery levels throughout urban commuting.
\end{itemize}

To the best of our knowledge, our work is one of few research papers that capture the aggregated impact of a new charging facility on urban mobility. Furthermore, the analysis is not only derived mathematically but also verified with real transport data from one of the most iconic metropolitan cities. The results are abstracted into functions that relevant organizations can use to plan the deployment of dynamic charging, which accelerates EV adoption. The table of notations used in our work is shown in Table~\ref{notation}.

\begin{table}
\centering
\captionsetup{font=normalsize}
\caption{Summary of notations}
\label{notation}
\begin{tabular}{p{1.2cm}| p{6.8cm}}
\hline
\textbf{Notation} & \textbf{Description} \\ \hline
$\lambda$           & density of the 1D Poisson Point Process \\
$r$                 & distance (in meters) from the city center \\
$r_{\rm min}$       & lowest value of $r$ at which the power law is obeyed \\
$\alpha$            & parameter of the power law function \\
$d_h$                 & horizontal distance between a source and a destination \\
$d_v$               & vertical distance between a source and a destination\\
$D_n$               & distance from a source to the nearest charging road\\
$\rho_c$               & trip portion (in percentage) travelled on charging roads \\
$e_c$               & the amount of energy charged in a trip \\
$D_\mathrm{N-HC}$          & distance from a source to the nearest horizontal charging road\\
$D_\mathrm{N-VC}$          & distance from a source to the nearest vertical charging road\\
$D_\mathrm{N-HNC}$         & distance from a source to the nearest horizontal non-charging road\\
$D_\mathrm{N-VNC}$         & distance from a source to the nearest vertical non-charging road\\
\hline

\end{tabular}
\end{table}


\section{Related Works}

In the literature, some previous studies have been introduced on the topic of dynamic charging system deployment~\cite{optimaldeploymentofcharginglanes,optimallocationofwirelesscharging,optimizationofstaticandynamic}. Most of those studies formulate this topic as optimization problems, in which specific components of the dynamic charging systems are optimized, e.g., maximizing the power received by EVs, minimizing infrastructure cost, and enhancing EV battery level~\cite{deploymentoptimizationofdynamic}. For example, a categorization and clustering approach to minimizing the total deployment cost while maintaining the state-of-charge for EVs is presented in~\cite{catcharger}. Analytical models to determine the optimum vehicle battery size and locations of power tracks in single-route and multi-route environments, which are particularly applicable to electric buses, using a Particle Swarm Optimization (PSO) algorithm, are illustrated in~\cite{optimalsystemdesign, systemoptimizationfordynamic}. A study on minimizing the capital costs of dynamic charging infrastructure while enabling EVs to travel among popular destinations in California is introduced in~\cite{wirelesschargingincali}.  In~\cite{optimalplacementroadnetworks}, the installation of wireless charging lanes is framed as an integer programming problem. Given a budget, the number of routes that benefit from the charging lanes is maximized. In~\cite{evroadgrid}, a charging path optimization algorithm is introduced to minimize the traveling time and the charging cost. However, these kinds of works do not capture the impact of dynamic charging system deployment on consumers. Besides dynamic charging, to facilitate vehicle electrification, some research on intelligent EV charging navigation~\cite{li2020effective} and other charging options have been introduced~\cite{li2018direct}. For instance, in~\cite{li2020efficient,li2019intelligent}, a direct vehicle-to-vehicle energy-exchange strategy has been explored. Unlike those studies, our research focuses specifically on how dynamic charging, a rising charging facility that simultaneously serves multiple EVs and mitigates charging stations' problems, benefits drivers in their daily trips and fosters EV adoption. 

\section{Dynamic Charging Deployment Strategy}
\label{deploymentstrategy}
In this section, we first review the goals of dynamic charging systems, i.e., charging roads, and then discuss two strategies for deploying dynamic charging in metropolitan cities. 

\subsection{Deployment Goals}
\label{deploymentgoals}
Dynamic charging systems are developed to alleviate the issues with charging stations, i.e., the long charging time and the range anxiety problem. Therefore, dynamic charging should be deployed to extend the driving range, make visits to charging stations less often, and reduce the need for more charging stations. Moreover, they need to ensure practicality by having high utilization, i.e., benefiting multiple cars simultaneously, and efficiency, i.e., meeting the charging demand using only a small number of charging roads. Based on these goals, we examine two deployment strategies in the following two subsections. 


\subsection{Baseline Strategy: Uniform Deployment}
A good starting point is to deploy dynamic charging systems uniformly across the city at a certain density. In other words, each road in the city will have an equal probability of being a charging road. This approach is simple yet may be sufficient for cities with uniformly distributed population and traffic. Moreover, it can serve as a baseline to compare against other well-planned strategies. The drawback of this strategy is that it may not be optimal for cities with non-uniformly distributed traffic and population. 

\subsection{Traffic-Based Deployment Strategy}

Since dynamic charging systems serve the people and their commutes, it may be beneficial to study the spatial distribution of population and traffic density, which are positively correlated. To this end, we refer to popular models in the literature, which found that in many urban cities, the population and traffic are often dense in the interior, and then sharply decline as we move to the outer suburbs. Indeed, in~\cite{simplespatialscaling}, it is revealed that the spatial distribution of active population, i.e., a mixture of working and residential population, the construction of road networks, and the socioeconomic interactions along roads all scale following a power law function in terms of the distance from the city center, expressed as

\begin{equation*}
    y \propto r^{-\alpha}, 
\end{equation*}
where $y$ can be the population, the road networks, or the socioeconomic interactions. $r$ is the distance from the city center, and $\alpha$ is a positive parameter. A high value of $\alpha$ indicates that the density falls sharply as the distance from the city center increases, while a low value of $\alpha$ signifies that the density declines more slowly. Despite its simplicity, the model demonstrates a good agreement with empirical data of several big cities such as Amsterdam, Beijing, Berlin, London, Los Angeles, Milan, and Tokyo~\cite{simplespatialscaling}. We also find that the model fits the traffic data in New York City, as described in Section~\ref{SpatialDistributionofUrbanTrips}.

The charging roads can therefore be deployed following a power law function from the city center. Compared to the uniform deployment strategy, this approach is more customized to fit the city traffic. Thus, it is likely to give better results for cities where traffic is non-uniform. However, the disadvantage is that the strategy is complex to develop and more complicated to analyze theoretically.

Having introduced the deployment goals and the two strategies, we proceed with the question of how to compare them. To this end, in the next section, we discuss the quantitative metrics that reflect the deployment goals stated in Section~\ref{deploymentgoals}.  

\section{Dynamic Charging Deployment Assessment}
\label{deploymentassesment}
\subsection{Quantitative Metrics}
As stated in Section~\ref{introduction}, our framework targets policy makers in either public institutions or private organizations who seek strategies on deploying charging roads in metropolitan cities. Thus, for this group of professionals, we propose two metrics that provide a high-level executive summary of the impact of charging road deployment on urban commuting trips. The metrics are defined below.

\begin{definition}[Probability distribution of the distance to the nearest charging road, i.e., $P(D_n < x)$]
It is the probability that the distance traveled from a given source to the nearest charging road is less than a positive number $x$. 
\end{definition}

\begin{definition}[Probability distribution of the trip portion traveled on charging roads, i.e., $P(\rho_c < x)$]
It is the probability that the trip portion (in percentage) traveled on charging road from a given source to a given destination is less than a positive number $x \in [0,100]$. 
\end{definition}

The first metric indicates how long a driver has to drive before meeting a charging road in a given trip, and the second describes how much the driver can benefit from the network of charging roads. These metrics are statistical, meaning that they should be understood as if the metrics were averaged over many trips in urban cities. Since the metrics capture traffic as a whole, they are not only useful for planning the deployment of charging roads, but also critical for comparing different deployment strategies as presented in Section~\ref{deploymentstrategy}. In addition, EV manufacturers can leverage the metrics to produce suitable battery sizes to match certain cities' road conditions.

\subsection{Analytical Framework \& System Model}
To compute the two metrics, we need a means to model the stochastic road networks in metropolitan cities, with the charging roads being deployed following a certain probability function. We select stochastic geometry as the primary tool for this analysis, since there are several reasons why stochastic geometry is well-suited for this problem. First and foremost, stochastic geometry is a statistical tool that models the road networks as stochastic processes. Thus, it lets us study the metrics as if we were averaging over all the dynamic sources and destinations of urban trips~\cite{stochasticgeometryforwireless,PLCPfoundations}. Second, it has been used extensively in the literature to model and characterize similar problems in transportation networks~\cite{geodesicsandflow,meantrafficbehavior,nguyen2020modeling,dynamicjointallocation,stochasticgeometryplanning,shortestpathdistance}.

Specifically, we model the road network as a homogeneous Manhattan Poisson Line Process (MPLP), which closely approximates the road networks in several metropolitan cities such as New York~\cite{gridasalgorithm}, Vancouver~\cite{dreamcityvancouver}, and Barcelona~\cite{citiesasemergentmodel}. A homogeneous Manhattan Poisson line process (MPLP) is a collection of random perpendicular lines in a 2D plane. These lines are generated from two Poisson point processes (PPP) with a density parameter $\lambda$ along the x-axis and y-axis, respectively. A PPP with density parameter $\lambda$ means on a line segment of length $d$, the number of points on that line segment follows a Poisson distribution with parameter $\lambda d$. 

The assignment of charging roads is then done using a thinning function, reflecting the charging probabilities of the roads. Thinning is a process of removing points from a point process. Each point has a probability of being kept or removed following a user-defined thinning function. In this research, the points are removed independently. For the uniform deployment strategy, each road has a probability $p$ of being equipped with dynamic charging. For the traffic-based deployment strategy, the thinning function is chosen to be a power law function that is defined as $g: \mathbb{R} \mapsto [0,1]$,

\begin{equation}
\label{gr}
    g(r) = 
    \begin{cases}
      (\frac{\mid r \mid}{r_{\rm min}})^{-\alpha} & \mid r \mid > r_{\rm min}\\
      1 & \mid r \mid \le r_{\rm min},
    \end{cases} 
\end{equation}
where $r_{\rm min}>0$ is some lowest value of $r$ at which the power law is obeyed. Intuitively, $g(r)$ represents the probability of equipping the road that is $r$ meters away from the city center with dynamic charging. Applying the thinning function on the road systems modeled as a homogeneous MPLP, we can model the system of charging roads as an inhomogeneous MPLP with intensity function $\lambda g(r)$. The system of non-charging roads is modeled as an inhomogeneous MPLP with intensity function $\lambda (1-g(r))$.

\subsection{Methodology Overview}

\begin{table*}[h]
\centering
\captionsetup{font=normalsize}
\caption{Events to calculate $D_n$, $\rho_c$}
\label{subevents}
\begin{tabular}{|c|c|c|}
\hline
Event                                                                                                                            & \begin{tabular}[c]{@{}c@{}}When the source road and \\ destination road are parallel\end{tabular} & \begin{tabular}[c]{@{}c@{}}When the source road \\ and destination road are perpendicular\end{tabular} \\ \hline
\begin{tabular}[c]{@{}c@{}}When both the source and the destination roads\\  are equipped with dynamic charging\end{tabular}     & Event $T_1$                                                                                       & Event $T_5$                                                                                            \\ \hline
\begin{tabular}[c]{@{}c@{}}When only the source road is equipped with \\ dynamic charging\end{tabular}                           & Event $T_2$                                                                                       & Event $T_6$                                                                                            \\ \hline
\begin{tabular}[c]{@{}c@{}}When only the destination road is equipped with \\ dynamic charging\end{tabular}                      & Event $T_3$                                                                                       & Event $T_7$                                                                                            \\ \hline
\begin{tabular}[c]{@{}c@{}}When both the source and the destination roads\\  are not equipped with dynamic charging\end{tabular} & Event $T_4$                                                                                       & Event $T_8$                                                                                            \\ \hline
\end{tabular}
\end{table*}


We derive the distribution of the distance to the nearest charging road, i.e., $D_n$, and the distribution of the portion traveled on charging roads, i.e., $\rho_c$, by breaking them down into eight events based on the locations of a driver as he or she makes the trip from a source to a destination, as presented in Table~\ref{subevents}.

Let $S$ and $D$ denote the source and the destination of a trip, respectively. The cumulative distribution functions (CDF) of $D_n$ and $\rho_c$ are given by
\begin{align}
&\mathbb{P}(D_n < x) = \sum_{i=1}^{8} \mathbb{P}(D_n < x|T_i)\mathbb{P}(T_i), \\
&\mathbb{P}(\rho_c < x) = \sum_{i=1}^{8} \mathbb{P}(\rho_c < x|T_i)\mathbb{P}(T_i).
\end{align}
To simplify the calculation process, we focus on the CDF of $D_n$ and $\rho_c$ given $S,D$ as follows:
\begin{align}
\label{eq1}
&\mathbb{P}(D_n < x|S,D) = \sum_{i=1}^{8} \mathbb{P}(D_n < x|T_i,S,D)\mathbb{P}(T_i|S,D), \\
\label{eq2}
&\mathbb{P}(\rho_c < x|S,D) = \sum_{i=1}^{8} \mathbb{P}(\rho_c < x|T_i,S,D)\mathbb{P}(T_i|S,D).
\end{align}

To compute the conditional probabilities $\mathbb{P}(D_n < x|T_i,S,D)$ and $\mathbb{P}(\rho_c < x|T_i,S,D)$, we need a consistent way to know how drivers go from a source to a destination. Therefore, we make an assumption about the driving behavior of drivers. That is, drivers will always choose the shortest route from a source to a destination. If there are several such routes, the drivers will choose the one that maximizes the time they spend on charging roads. More details on the computation of equations~\eqref{eq1} and~\eqref{eq2} are given in the following subsections.

\subsection{A Divide-and-Conquer Strategy}

We adopt a `divide-and-conquer' approach for all cases of $T$, i.e., further breaking each event $T_i$ into smaller subcases. The same methodology is applied to each event $T_i$. Thus, to avoid repetitiveness, we hereby show a representative example of event $T_3$.

\begin{figure}[h]
\centering
\includegraphics[width=1\columnwidth]{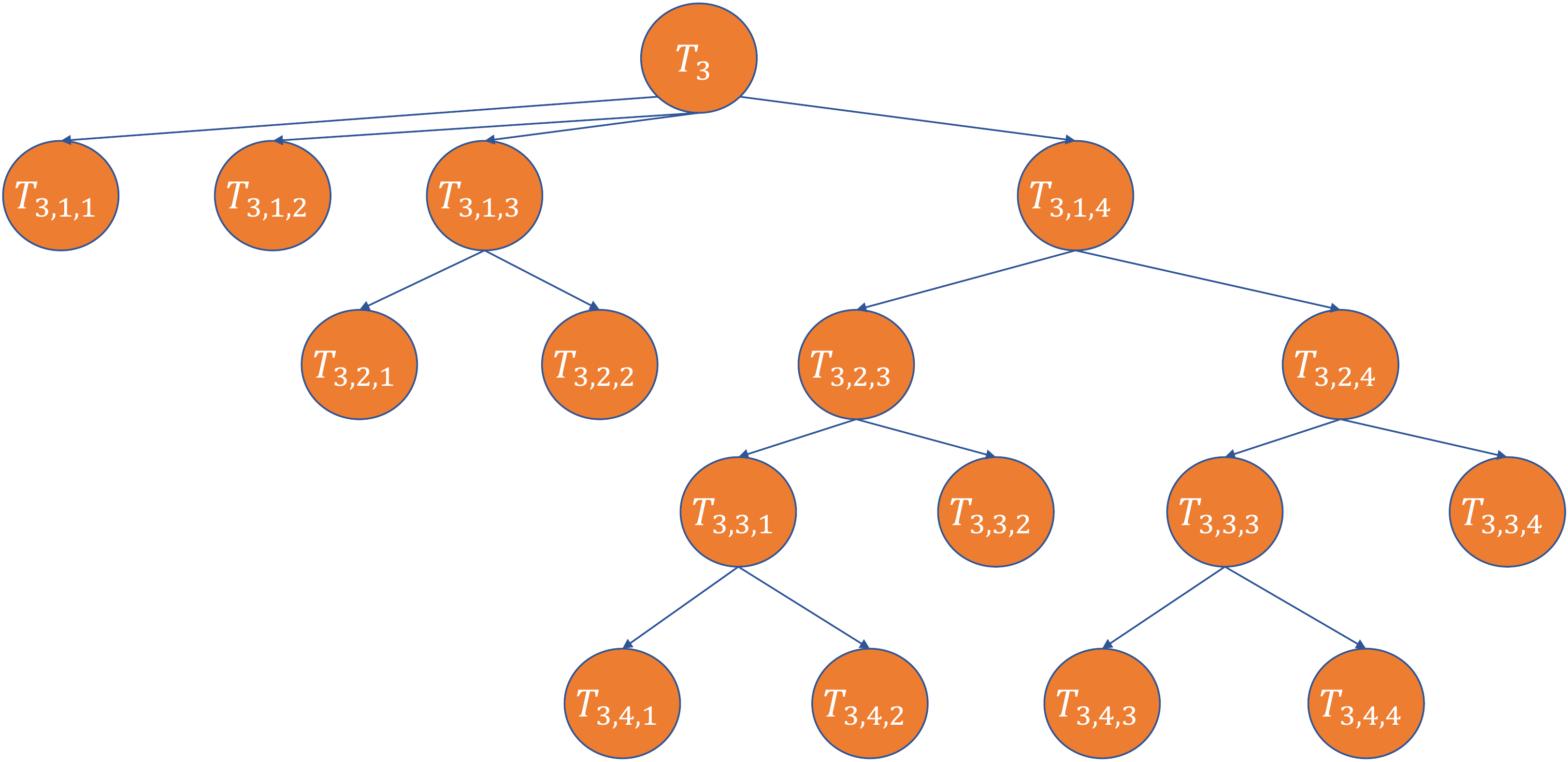}
\caption{Event $T_3$: when the source road and the destination road are parallel and only the destination road is equipped with dynamic charging.}
\label{E3}
\end{figure}

To compute $\mathbb{P}(D_n < x|T_3,S,D)$, we further divide it into subcases in a probability tree, as shown is Fig.~\ref{E3}.

\begin{figure}[h]
\centering
\captionsetup[subfigure]{font=scriptsize,labelfont=normalsize}
\subfloat[Node $T_{3,1,1}$\label{E311}]{%
  \includegraphics[width=0.33\columnwidth,keepaspectratio]{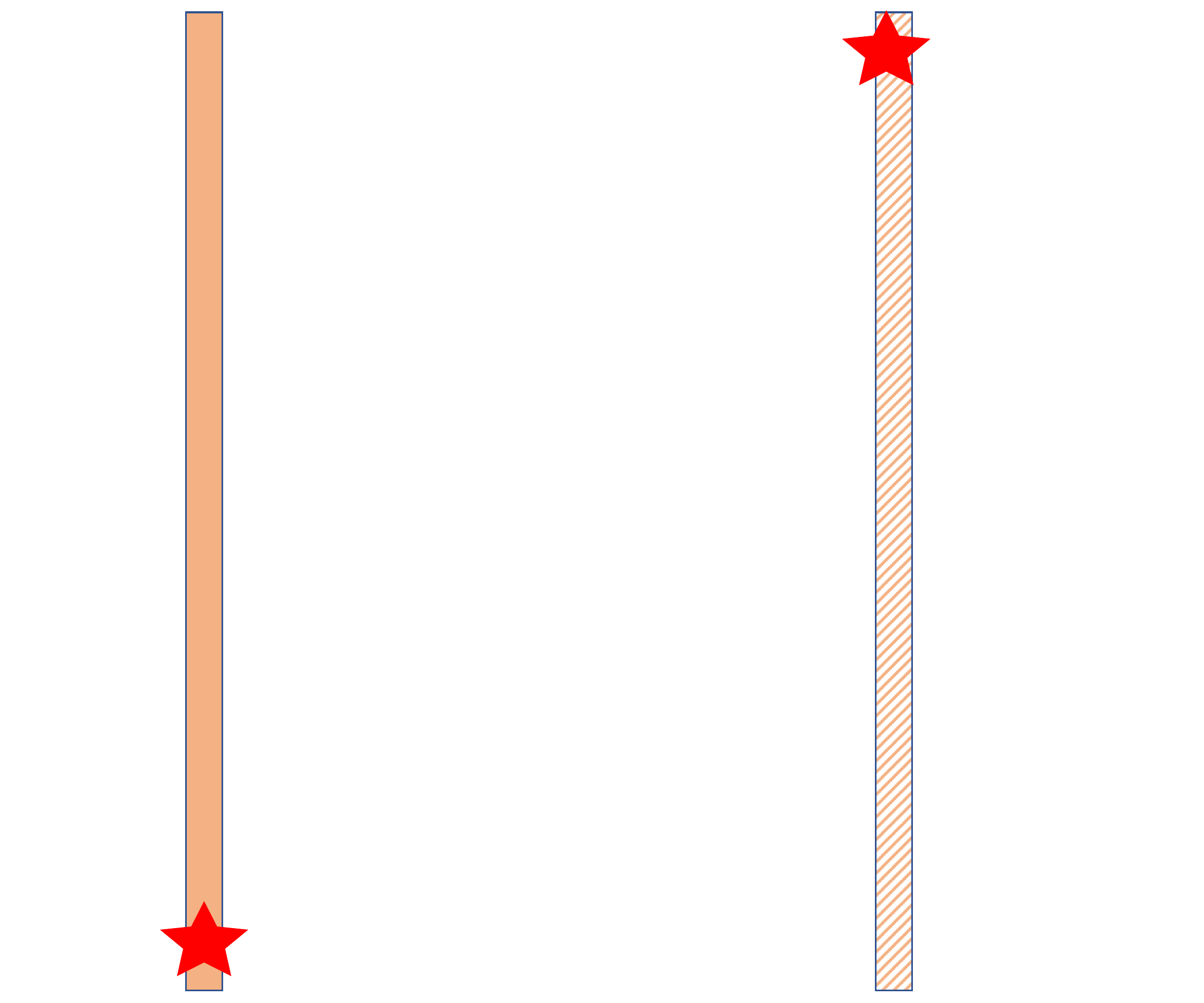}%
}\hfill
\subfloat[Node $T_{3,1,2}$\label{E312}]{%
  \includegraphics[width=0.33\columnwidth,keepaspectratio]{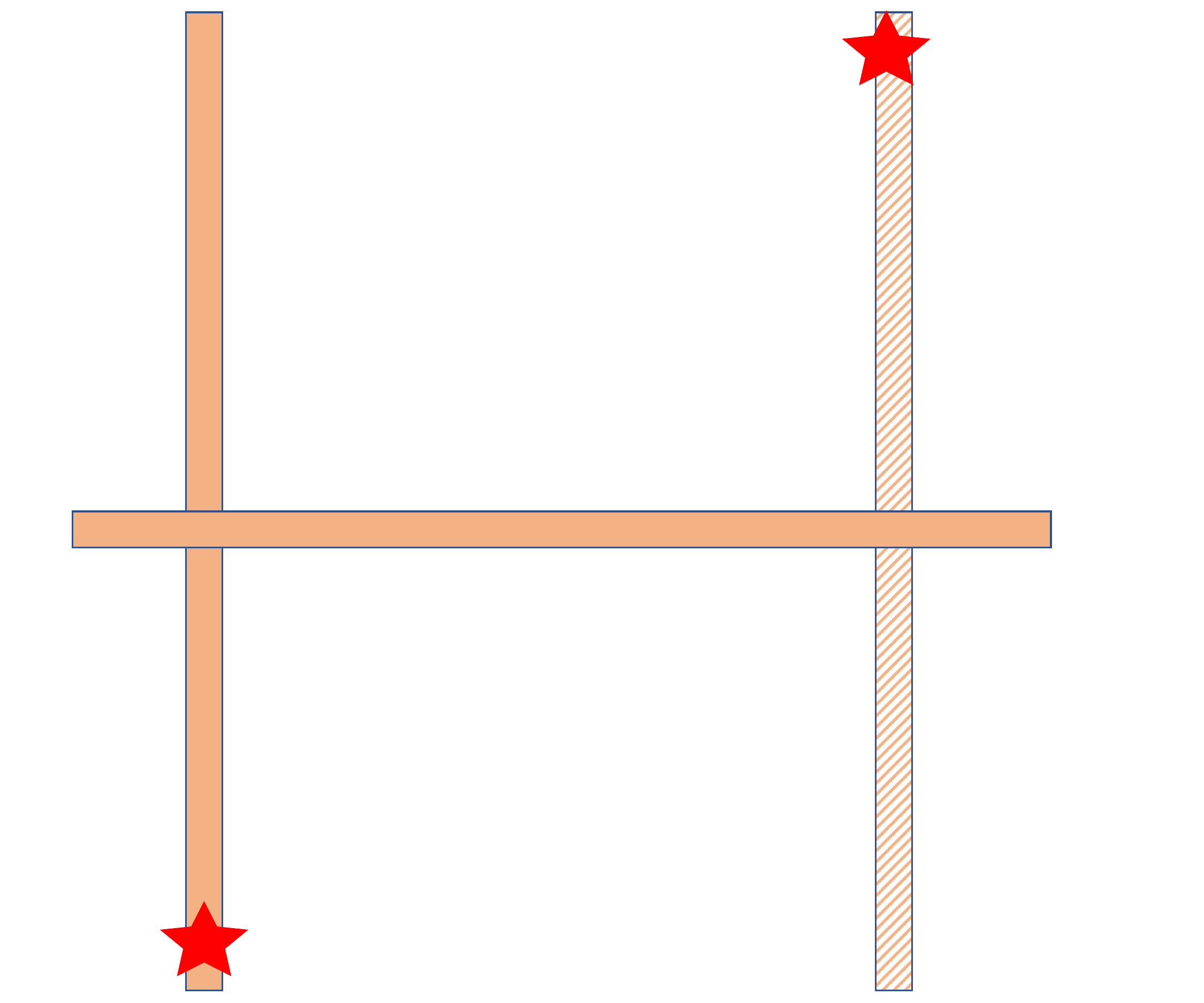}%
}
\hfill
\subfloat[Node $T_{3,2,1}$\label{E321}]{%
  \includegraphics[width=0.33\columnwidth,keepaspectratio]{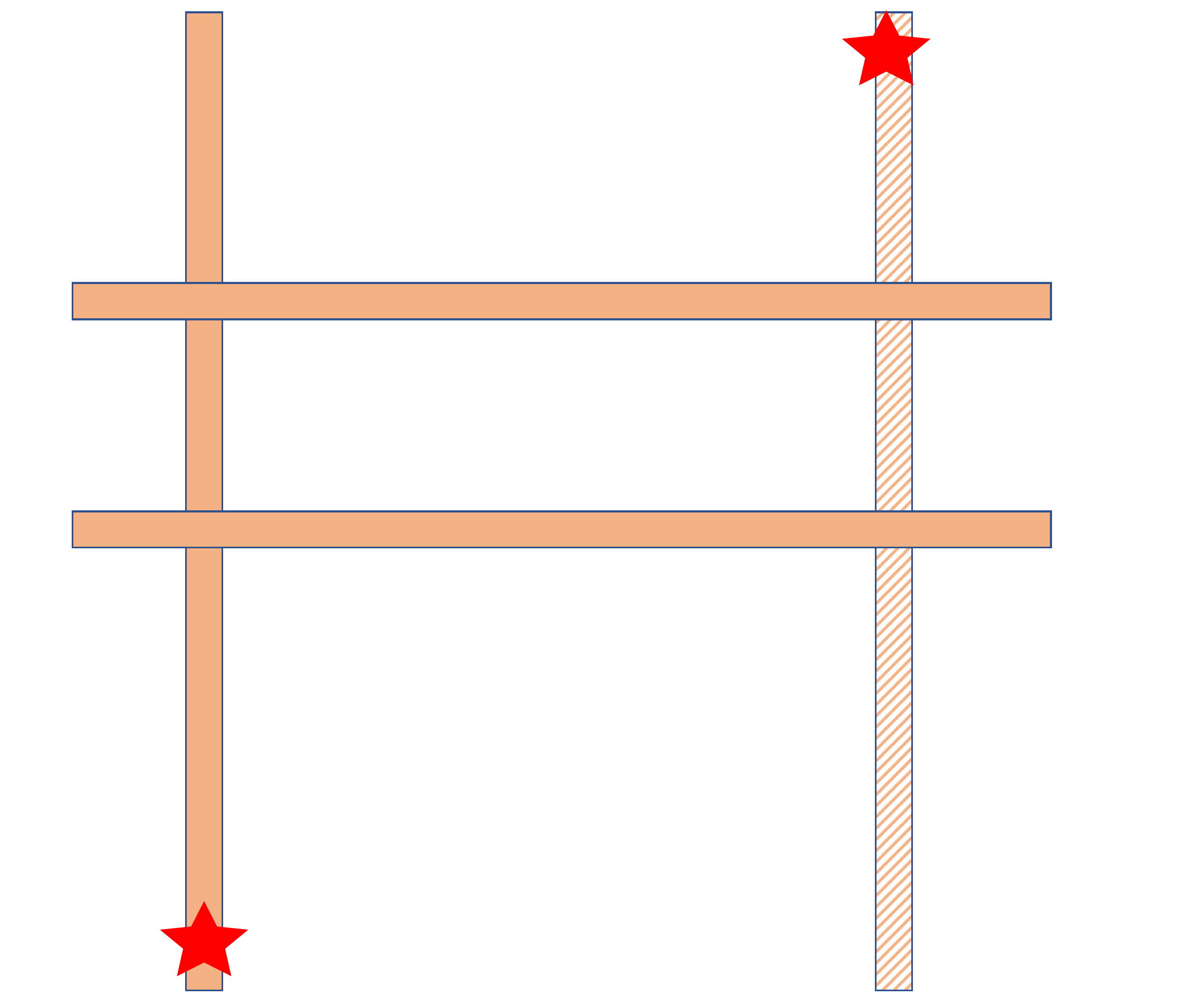}%
}

\subfloat[Node $T_{3,2,2}$\label{E322}]{%
  \includegraphics[width=0.33\columnwidth,keepaspectratio]{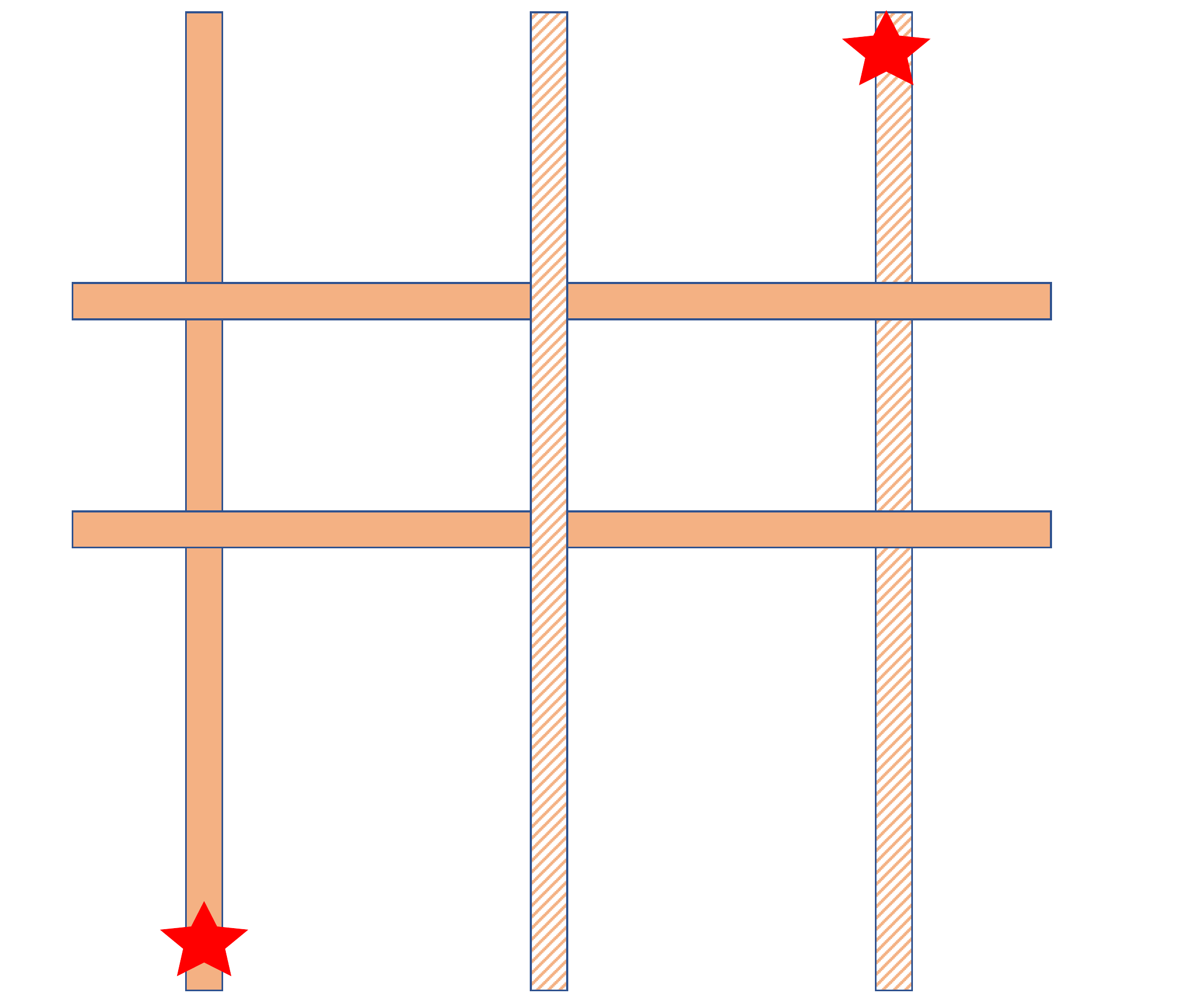}%
}\hfill
\subfloat[Node $T_{3,3,1}$\label{E331}]{%
  \includegraphics[width=0.33\columnwidth,keepaspectratio]{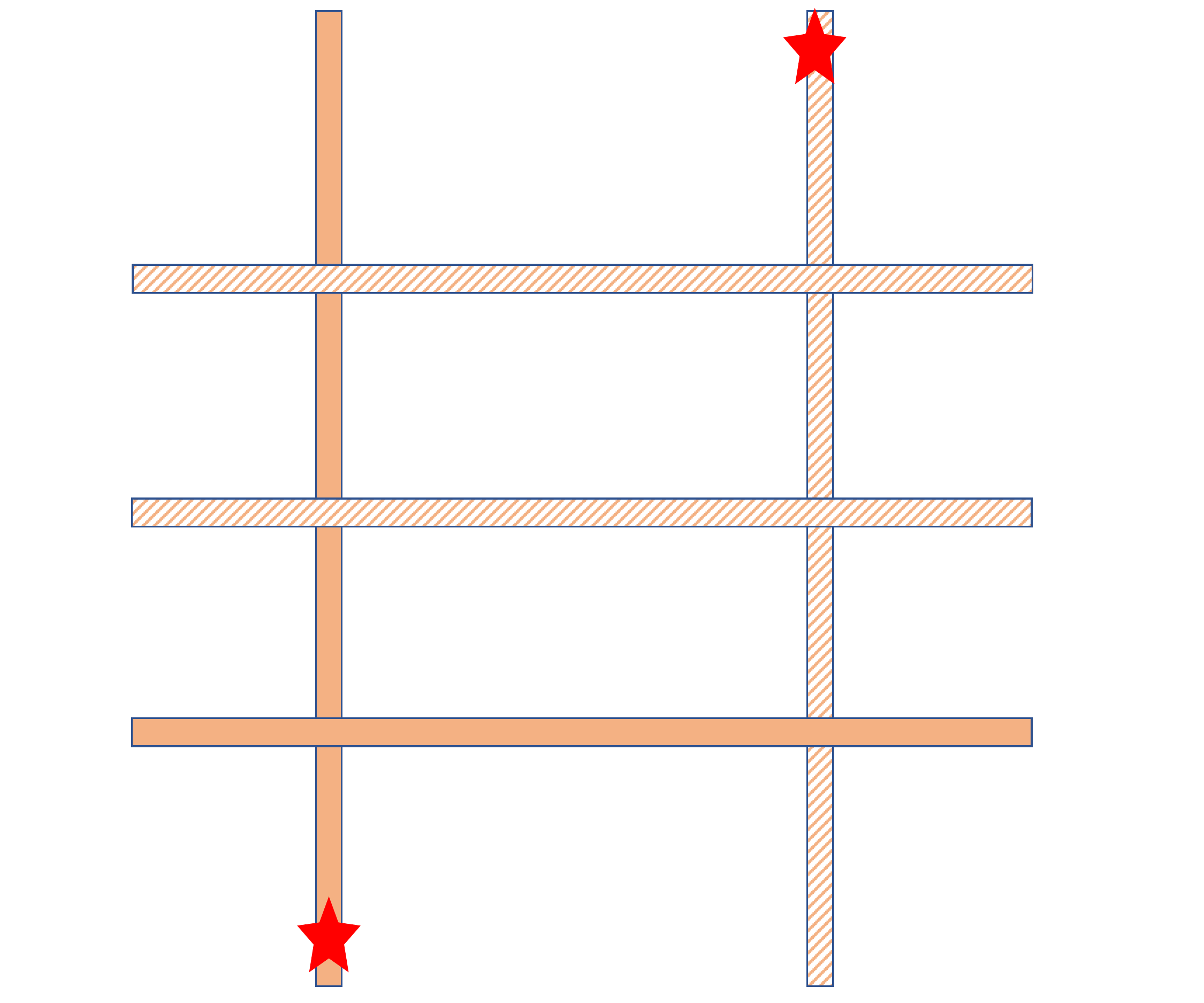}%
}
\hfill
\subfloat[Node $T_{3,3,2}$\label{E332}]{%
  \includegraphics[width=0.33\columnwidth,keepaspectratio]{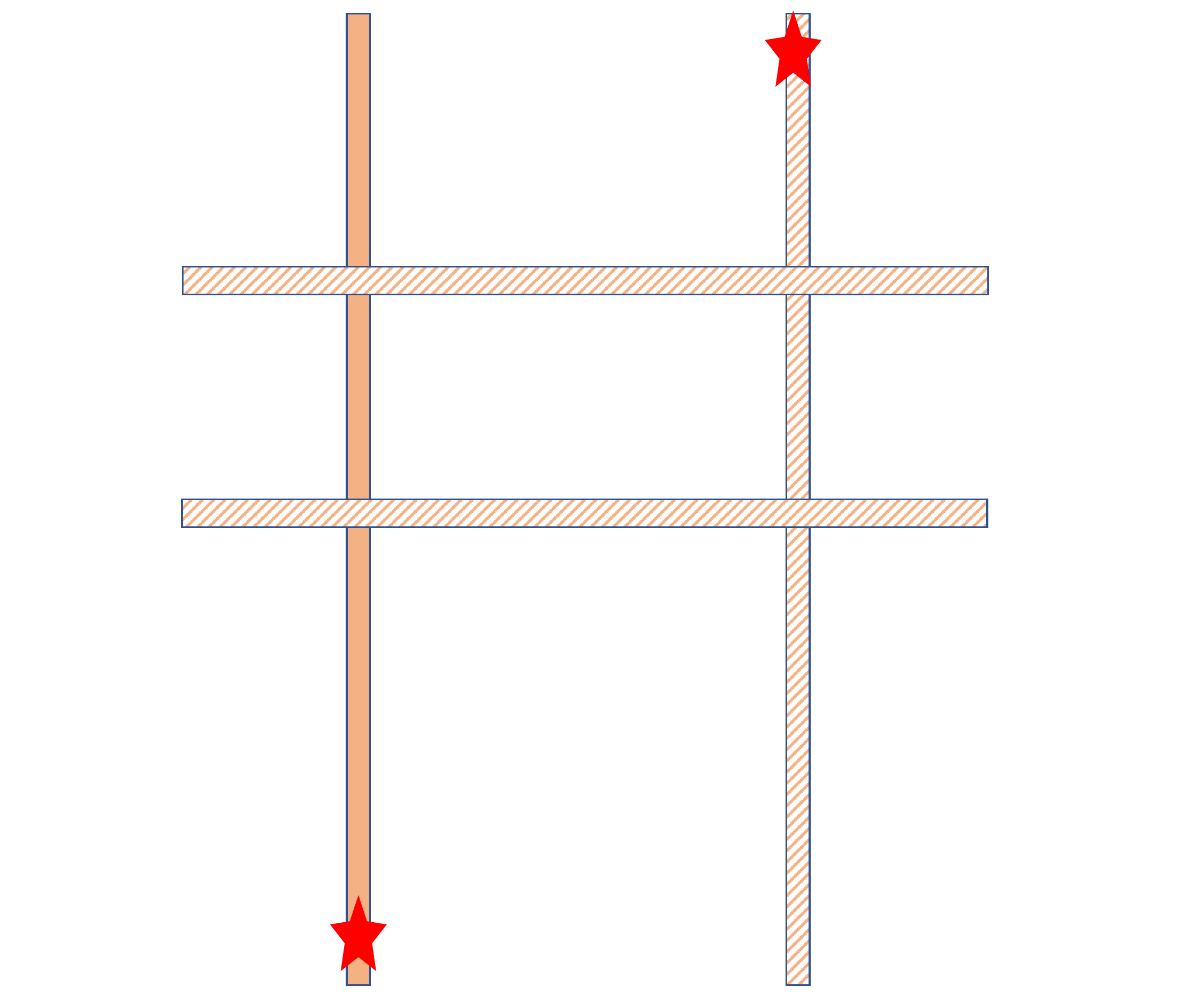}%
}

\subfloat[Node $T_{3,3,3}$\label{E333}]{%
  \includegraphics[width=0.33\columnwidth,keepaspectratio]{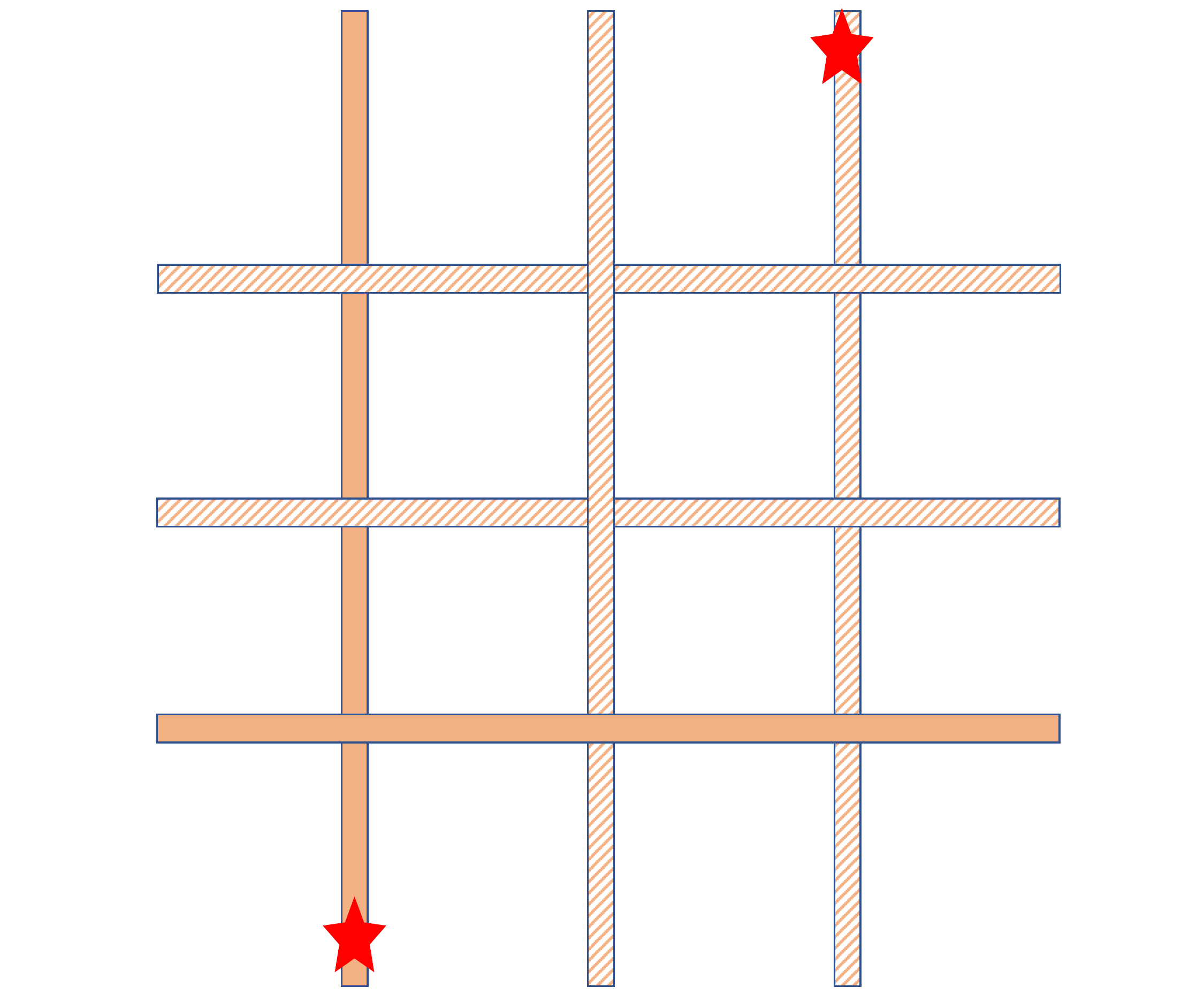}%
}\hfill
\subfloat[Node $T_{3,3,4}$\label{E334}]{%
  \includegraphics[width=0.33\columnwidth,keepaspectratio]{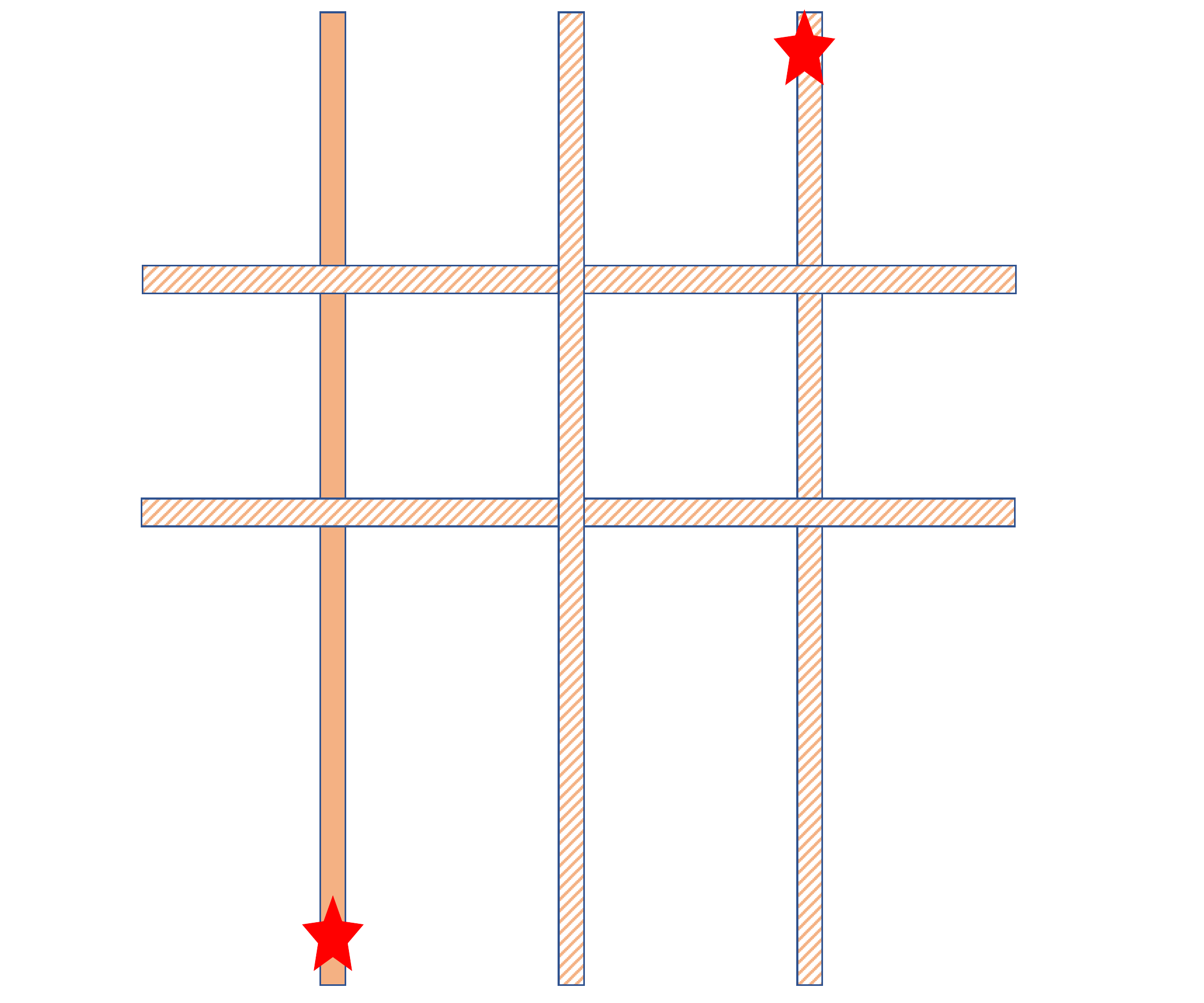}%
}\hfill
\subfloat{%
  \includegraphics[width=0.33\columnwidth,keepaspectratio]{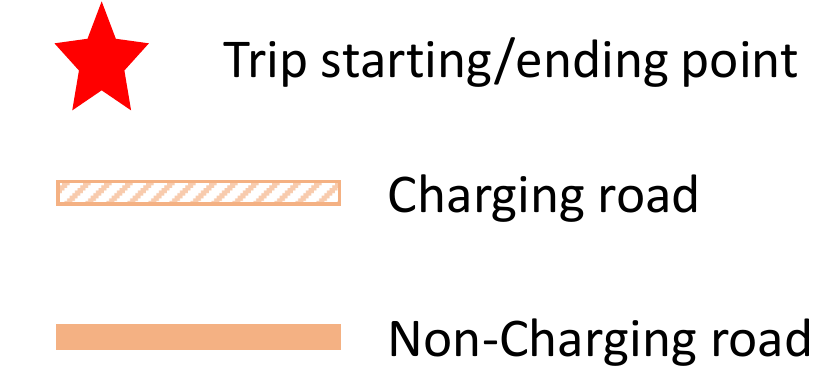}%
}

\caption{An illustration of the nodes of probability tree $T_3$.}
\label{e3sub}
\end{figure}

The notations for the subevents or nodes are three-fold. The first subscript denotes the event as in Table~\ref{subevents}. The second subscript means the level of depth in the probability tree, as shown in Fig.~\ref{E3}. The third subscript represents the index of the event at that level. A demonstration for those subevents of event $T_3$ is presented in Fig.~\ref{e3sub}. Consequently, the distribution of $D_n$ given $T_3$, $S$, and $D$ can be derived as follows:

\begin{align}
&\mathbb{P}(D_n < x | T_3,S,D)\mathbb{P}(T_3,S,D)\nonumber \\& = \sum_{i=1}^{10} \mathbb{P}(D_n < x | L_{3,i},S,D)\mathbb{P}(L_{3,i}|S,D),\label{eqL}
\end{align} 
where $L_{3,i}$'s are the intersection of consecutive events from the root to the leaves of tree $T_3$. For example, $L_{3,1} = T_{3,1,1} \cap T_{3}$, $L_{3,2} = T_{3,1,2} \cap T_{3}$, $L_{3,3} = T_{3,2,1} \cap T_{3,1,3} \cap T_{3}$, and so on (See Fig.~\ref{E3}).

The details of nodes in the tree $T_3$ are listed below. Each node corresponds to a case that drivers can face when traveling from a source to a destination.
\begin{itemize}[wide, labelwidth=!, labelindent=0pt]
    \item Node $T_{3,1,1}$: when drivers travel through no horizontal roads, as presented in Fig. \ref{E311}; 
    \item Node $T_{3,1,2}$: when drivers travel through only one horizontal road that is not equipped with dynamic charging, as presented in Fig. \ref{E312}; 
    \item Node $T_{3,1,3}$: when the route from the source to the destination has at least two horizontal roads that are all not equipped with dynamic charging;
    \begin{itemize}[wide, labelwidth=!, labelindent=0pt]
        \item Node $T_{3,2,1}$: when drivers pass through no vertical roads that are equipped with dynamic charging, as presented in Fig. \ref{E321};
        \item  Node $T_{3,2,2}$: when drivers pass through at least one vertical road that is equipped with dynamic charging, as presented in Fig. \ref{E322}; 
    \end{itemize}
    \item Node $T_{3,1,4}$: when the route from the source to the destination has at least one horizontal road that is equipped with dynamic charging;
    \begin{itemize}[wide, labelwidth=!, labelindent=0pt]
        \item Node $T_{3,2,3}$: when drivers travel through no vertical roads that are equipped with dynamic charging;
        \begin{itemize}[wide, labelwidth=!, labelindent=0pt]
            \item Node $T_{3,3,1}$: when the nearest horizontal road from the source is not equipped with dynamic charging, as presented in Fig. \ref{E331};
            \begin{itemize}[wide, labelwidth=!, labelindent=0pt]
                \item Node $T_{3,4,1}$: when drivers decide to take the nearest horizontal road that is equipped with dynamic charging; \item Node $T_{3,4,2}$: when drivers take the nearest horizontal road that is not equipped with dynamic charging;
            \end{itemize}
            \item Node $T_{3,3,2}$: when the nearest horizontal road from the source is equipped with dynamic charging, as presented in Fig. \ref{E332}; 
        \end{itemize}
        \item Node $T_{3,2,4}$: when drivers pass through at least one vertical road that is equipped with dynamic charging;
        \begin{itemize}[wide, labelwidth=!, labelindent=0pt]
            \item Node $T_{3,3,3}$: when the nearest horizontal road from the source is not equipped with dynamic charging, as presented in Fig. \ref{E333};
            \begin{itemize}[wide, labelwidth=!, labelindent=0pt]
                \item Node $T_{3,4,3}$: when drivers decide to take the nearest vertical road that is equipped with dynamic charging;
                \item Node $T_{3,4,4}$: when drivers decide to take the nearest horizontal road that is equipped with dynamic charging;
            \end{itemize}
            \item Node $T_{3,3,4}$: when the nearest horizontal road from the source is equipped with dynamic charging, as presented in Fig. \ref{E334}; 
        \end{itemize}
    \end{itemize}
\end{itemize}
To compute each term in equation~\eqref{eqL}, in section~\ref{summarydistr}, we first derive some prerequisite propositions that will be used repeatedly in later proof. Then, in section~\ref{calDn}, we show a detailed calculation of a subcase of event $T_3$, i.e., a term in equation~\eqref{eqL}, to illustrate our idea.

\subsection{Summary of important distributions}
\label{summarydistr}

\begin{prop}
\label{propDnvc}
Let $D_\mathrm{N-VC}$ be the Manhattan distance from the source to the nearest vertical road that is equipped with dynamic charging. The CDF of $D_\mathrm{N-VC}$ is 
\begin{align*}
    &F_{D_\mathrm{N-VC}}(x) =  \mathbb{P}(D_\mathrm{N-VC} < x) \\&= 1-2\int_{-\infty}^{\infty} \int_{-\infty}^{s} e^{-\lambda \int_{s-x}^{s} g(r) {\rm d}r} f_D(d)  f_S(s)  {\rm d}d  {\rm d}s.
\end{align*}
\end{prop}
\begin{IEEEproof}
See Appendix~\ref{proofdvc}.
\end{IEEEproof}

\begin{prop}
\label{propDnvcSD}
The CDF of $D_\mathrm{N-VC}|S,D$ is 
\begin{align*}
    &F_{D_\mathrm{N-VC}|S,D}(x) =
    \mathbb{P}(D_\mathrm{N-VC} < x|S,D) = 1- e^{-\lambda \int_{A} g(r) {\rm d}r}, 
\end{align*}
where A is a segment of length $x$ from $S$, i.e., $A = (s, s+x)$ if $S<D$ or $A = (s-x, s)$ if $S>D$. 

The PDF $D_\mathrm{N-VC}|S,D$ is
\begin{align*}
    &f_{D_\mathrm{N-VC}|S,D}(x|s,d, s<d) = \frac{{\rm d}}{{\rm d}x} F_{D_\mathrm{N-VC}|S,D}(x)
    \\&= \lambda g(s+x) e^{-\lambda \int_{s}^{s+x} g(r) {\rm d}r}, \\
    &f_{D_\mathrm{N-VC}|S,D}(x|s,d, s>d) = \frac{{\rm d}}{{\rm d}x} F_{D_\mathrm{N-VC}|S,D}(x)
    \\&= \lambda g(s-x) e^{-\lambda \int_{s-x}^{s} g(r) {\rm d}r}.
\end{align*}
\end{prop}

The closed form for $F_{D_\mathrm{N-VC}|S,D}(x)$, which means the closed form for $\int_{A} g(r) {\rm d}r$, is derived in Appendix~\ref{grclosedform}. 

\begin{prop}
The CDF and PDF of $D_\mathrm{N-HC}|S,D$ are the same as those of $D_\mathrm{N-VC}|S,D$.
\end{prop}
\begin{IEEEproof}
The proof of $F_{D_\mathrm{N-HC}|S,D}(x)$ and $f_{D_\mathrm{N-HC}|S,D}(x)$ are similar to those of $F_{D_\mathrm{N-VC}|S,D}(x)$ and $f_{D_\mathrm{N-VC}|S,D}(x)$ given in Proposition~\ref{propDnvcSD}.
\end{IEEEproof}

\begin{prop}
\label{propDvnc}
Let $D_\mathrm{N-VNC}$ be the Manhattan distance from the source to the nearest vertical road that is not equipped with dynamic charging. The CDF of $D_\mathrm{N-VNC}$ is 
\begin{align*}
    &\mathbb{P}(D_\mathrm{N-VNC} < x) \\&=  1-2\int_{-\infty}^{\infty} \int_{-\infty}^{s} e^{-\lambda \int_{s-x}^{s} (1-g(r)) {\rm d}r} f_D(d)  f_S(s)  {\rm d}d  {\rm d}s.
\end{align*}
\end{prop}
\begin{IEEEproof}
The proof of $\mathbb{P}(D_\mathrm{N-VNC} < x)$ is similar to that of $\mathbb{P}(D_\mathrm{N-VC} < x)$ given in Proposition~\ref{propDnvc}.
\end{IEEEproof}

\begin{prop}
\label{propDvncSD}
The CDF of $D_\mathrm{N-VNC}|S,D$ is 
\begin{align*}
    &\mathbb{P}(D_\mathrm{N-VNC} < x|S,D, S <D) = 1- e^{-\lambda \int_{s}^{s+x} (1-g(r)) {\rm d}r}, \\&
    \mathbb{P}(D_\mathrm{N-VNC} < x|S,D, S > D) = 1- e^{-\lambda \int_{s-x}^{s} (1-g(r)) {\rm d}r}.
\end{align*}

The PDF $D_\mathrm{N-VNC}|S,D$ is
\begin{align*}
    &f_{D_\mathrm{N-VNC}|S,D}(x|s,d,s<d) = \frac{{\rm d}}{{\rm d}x} F_{D_\mathrm{N-VNC}|S,D}(x)
     \\&=\lambda (1-g(s+x)) e^{-\lambda \int_{s}^{s+x} (1-g(r)) {\rm d}r}, \\
    &f_{D_\mathrm{N-VNC}|S,D}(x|s,d,s>d) = \frac{{\rm d}}{{\rm d}x} F_{D_\mathrm{N-VNC}|S,D}(x)
     \\&=\lambda (1-g(s-x)) e^{-\lambda \int_{s-x}^{s} (1-g(r)) {\rm d}r}.
\end{align*}
\end{prop}

\begin{prop}
The CDF and PDF of $D_\mathrm{N-HNC}|S,D$ are the same as those of $D_\mathrm{N-VNC}|S,D$ given in Proposition~\ref{propDvncSD}.
\end{prop}

\begin{prop}
\label{propX1}
Let $d_h$ be the horizontal distance between $S$ and $D$. Let $X_1$ be the distance from the source's nearest vertical road that is not equipped with dynamic charging to the next vertical road equipped with dynamic charging, provided that both road types are present on the route from the source to the destination. The CDF of $X_1$ given S and D is
\begin{align*}
    &\mathbb{P}(X_1 < x|S,D, S<D) \\&=\frac{\int_{x}^{d_h} (e^{-\lambda \int_{s}^{s+t-x} (1-g(r)) {\rm d}r} - e^{-\lambda \int_{s}^{s+t} (1-g(r)) {\rm d}r})}{\int_{0}^{d_h} F_{D_\mathrm{N-VNC}|S,D}(t)} 
    \\&\times\frac{f_{D_\mathrm{N-VC}|S,D}(t|s,d,s<d) {\rm d}t}{ f_{D_\mathrm{N-VC}|S,D}(t|s,d,s<d){\rm d}t}.
\end{align*}
\end{prop}
\begin{IEEEproof}
See Appendix~\ref{proofx1}.
\end{IEEEproof}

\begin{prop}
Let $d_v$ be the vertical distance between $S$ and $D$. Let $X_2$ be the distance from the source's nearest horizontal road that is not equipped with dynamic charging to the next horizontal road equipped with dynamic charging, provided that both road types are present on the route from the source to the destination. The CDF of $X_2$ given S and D is
\begin{align*}
    &\mathbb{P}(X_1 < x|S,D, S<D) \\&= \frac{\int_{x}^{d_v} (e^{-\lambda \int_{s}^{s+t-x} (1-g(r)) {\rm d}r} - e^{-\lambda \int_{s}^{s+t} (1-g(r)) {\rm d}r}) }{\int_{0}^{d_v} F_{D_\mathrm{N-VNC}|S,D}(t) } 
    \\&\times\frac{f_{D_\mathrm{N-VC}|S,D}(t|s,d,s<d) {\rm d}t}{f_{D_\mathrm{N-VC}|S,D}(t|s,d,s<d){\rm d}t}.
\end{align*}
\end{prop}
\begin{IEEEproof}
The proof for $\mathbb{P}(X_1 < x|S,D, S<D)$ is similar to that of $\mathbb{P}(X_1 < x|S,D, S<D)$ given in Proposition~\ref{propX1}.
\end{IEEEproof}

\begin{prop}
Let $D_\mathrm{N-HNC}$ be the Manhattan distance from the source to the nearest vertical road that is not equipped with dynamic charging. The CDF of $D_\mathrm{N-HNC}$ is the same as that of $D_\mathrm{N-VNC}$ given in Proposition~\ref{propDvnc}. 
\end{prop}

\begin{prop}
\label{propDnhc}
Let $D_\mathrm{N-HC}$ be the Manhattan distance from the source to the nearest horizontal road that is equipped with dynamic charging. The CDF of $D_\mathrm{N-HC}$ is the same as that of $D_\mathrm{N-VC}$.
\end{prop}
\begin{IEEEproof}
The proof of $\mathbb{P}(D_\mathrm{N-HC} < x)$ is similar to that of $\mathbb{P}(D_\mathrm{N-VC} < x)$ given in Proposition~\ref{propDnvc}.
\end{IEEEproof}

\subsection{Calculation of $\rho_c$ and $D_n$}
\label{calDn}
To complete the computation in equation~\eqref{eqL}, we apply the same methodology to each of its terms. To avoid repetitiveness, we take the leaf ending at event $T_{3,4,1}$ as an example. Since this is the fifth leaf of tree $T_3$, event $L_{3,5} = T_{3,4,1} \cap T_{3,3,1} \cap T_{3,2,3} \cap T_{3,1,4} \cap T_3$ (see Fig.~\ref{E3}). Probability
\begin{align*}
                &\mathbb{P}(L_{3,5}|S,D)=\mathbb{P}(T_{3,4,1}|T_{3,3,1},T_{3,2,3},T_{3,1,4},T_3,S,D)
                \times\\&\mathbb{P}(T_{3,3,1}|T_{3,2,3},T_{3,1,4},T_3,S,D)
                \mathbb{P}(T_{3,2,3}|T_{3,1,4},T_3,S,D)
                \times\\&\mathbb{P}(T_{3,1,4}|T_3,S,D)\mathbb{P}(T_3|S,D) .
\end{align*}

Since $T_3$ denotes the event when the source and destination roads are parallel and only the destination road is equipped with dynamic charging, the probability of event $T_3|S,D$ is $(1-g(s))g(d)$, where $s$, $d$ are the distances from the source and the destination to the city center, respectively.

Assume that $S<D$, the computation of $\mathbb{P}(L_{3,5}|S,D)$, i.e., component probabilities are listed below.
\begin{align*}
    &\mathbb{P}(T_3|S,D) = (1-g(s))g(d),
    \\&\mathbb{P}(T_{3,1,4}|T_3,S,D) = 1 - e^{-\lambda \int_{s}^{s+d_v} g(r) {\rm d}r},
    \\&\mathbb{P}(T_{3,2,3}|T_{3,1,4},T_3,S,D) = e^{-\lambda \int_{s}^{s+d_h} g(r) {\rm d}r},
    \\&\mathbb{P}(T_{3,3,1}|T_{3,2,3},T_{3,1,4},T_3,S,D) \\&= \mathbb{E}_{D_\mathrm{N-HC}}[\mathbb{P}(T_{3,3,1}|T_{3,2,3},T_{3,1,4},T_3,D_\mathrm{N-HC},S,D)] 
    \\&= \mathbb{E}_{D_\mathrm{N-HC}}[1- e^{-\lambda \int_{s}^{s+D_\mathrm{N-HC}} 1-g(r) {\rm d}r}]
     \\&= \int_{0}^{d_v} (1-e^{-\lambda \int_{s}^{s+a} 1-g(r) {\rm d}r})\frac{f_{D_\mathrm{N-HC}(a)}}{F_{D_\mathrm{N-HC}}(d_v)}{\rm d}a,
     \\&\mathbb{P}(T_{3,4,1}|T_{3,3,1},T_{3,2,3},T_{3,1,4},T_3,S,D) = F_{X_2}(d_h).
\end{align*}

After calculating $\mathbb{P}(L_{3,5}|S,D)$, we move on to $\mathbb{P}(D_n < x|L_{3,5},S,D)$ and $\mathbb{P}(\rho_c < x|L_{3,5},S,D)$ as follows: 
\begin{align*}
     &\mathbb{P}(D_n < x|L_{3,5},S,D) = \Psi_1 (s,d \lambda, \alpha, r_{\rm min}, d_h, d_v, x),
     \\&\mathbb{P}(\rho_c < x|L_{3,5},S,D) = \Psi_2 (s,d, \lambda, \alpha, r_{\rm min}, d_h, d_v, x), 
\end{align*}
where $\Psi_1()$ and $\Psi_2()$ are the functions of the (charging) road density (i.e., $\lambda, \alpha, r_{\rm min}$) and the trip detail (i.e., $s,d,d_h, d_v, x$). The full forms and proof of $\Psi_1()$ and $\Psi_2()$ are given in Appendix~\ref{proofDnDc}.

\section{Case Studies}
\label{casestudies}
\subsection{Datasets}
We collect datasets on actual traffic in two urban cities: New York City (USA) and Xi'an (China). The New York City (NYC) dataset was released by the New York City Taxi and Limousine Commission (TLC)~\cite{tlc}. It consists of over 33 million trip records of yellow taxis over a 6-month period, from July 2019 to December 2019. Yellow taxis are the iconic taxis of New York and are the only serviced vehicles permitted to respond to street hails from passengers in all five boroughs of New York City. From the trip records, we mainly utilize the locations of the pickup and drop-off points. These locations are numbered as integers in the range of 1 to 263, corresponding to 263 taxi zones roughly based on the NYC Department of City Planning's Neighborhood Tabulation Areas. Since this dataset provides traffic information across NYC in neighborhood-level resolution, it is suitable to develop city-wide strategies for the deployment of charging roads. 

The second traffic dataset comes from Didi Chuxing Technology Co., which is a Chinese vehicle-for-hire company with over five hundred million users~\cite{didi}. The dataset contains trip records in an area of about 50 square kilometers inside the third ring road of Xi'an in October 2016. Unlike the NYC dataset, the Xi'an dataset zooms into an area with dense population and traffic~\cite{spatiotemporal}. Therefore, it is best used to examine intra-neighborhood deployment strategies that are locally optimized for populated regions of urban cities. 

For the existing road networks of NYC and Xi'an, we utilize the third dataset, which OpenStreetMap provides via the package OSMnx~\cite{osmnx}. In this dataset, the driving streets in those cities are represented as graphs whose nodes are the road intersections or deadends, and edges are the road portions connecting the nodes.  

\subsection{Spatial Distribution of Urban Trips}
\label{SpatialDistributionofUrbanTrips}
\begin{figure}[h]
\centering
\captionsetup[subfigure]{font=footnotesize}
\includegraphics[width=0.9\linewidth,keepaspectratio]{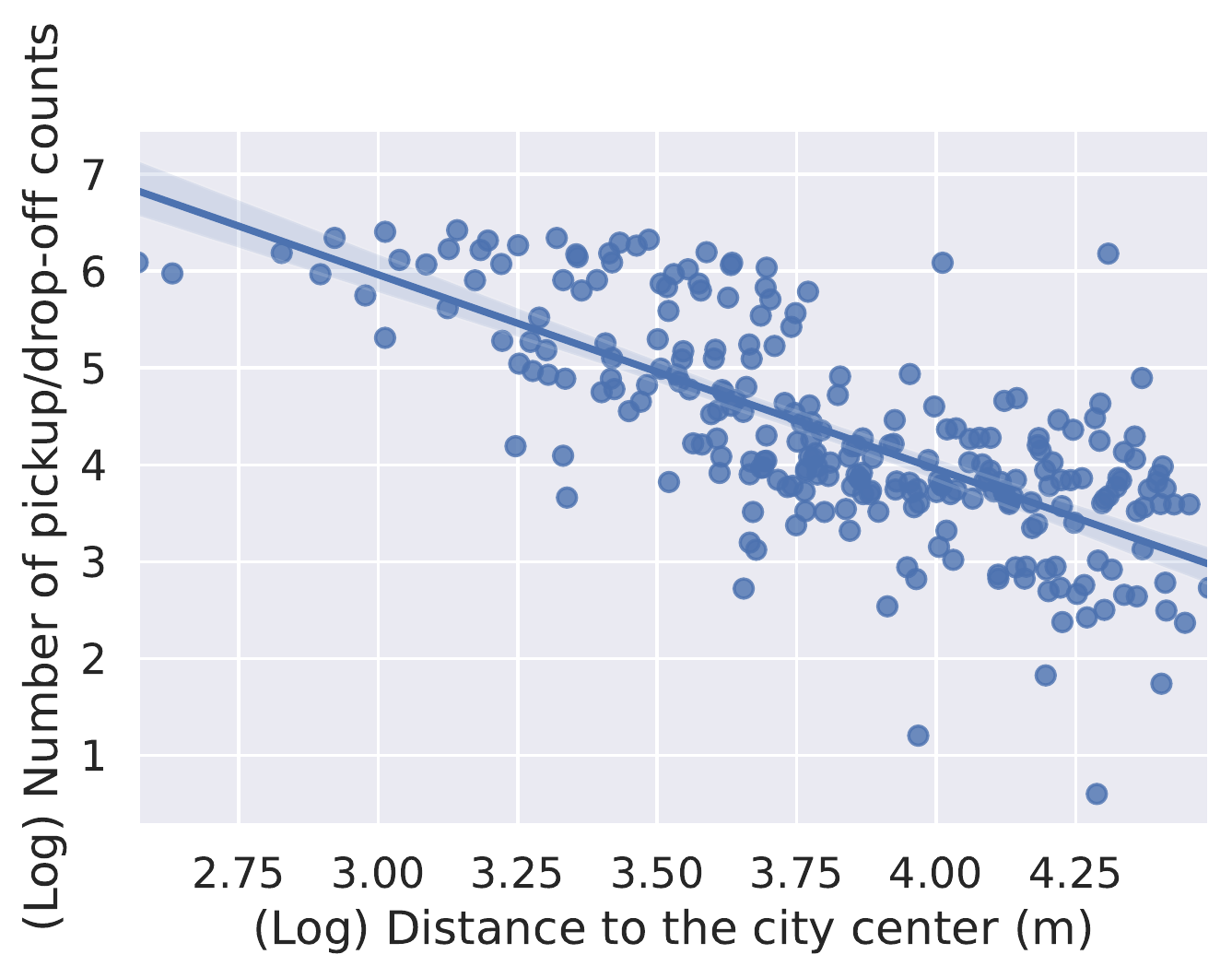}
\caption{Scatter plot of the taxi zones in New York City fitted with a linear regression model and 95\% confidence interval.}
\label{taxilr}
\end{figure}

This subsection mainly focuses on the NYC dataset, since it contains trips across the whole New York City. To investigate the spatial distribution of the trips, we aggregate the pickup/drop-off counts of each taxi zones and notice that zones with a large number of pickup/drop-offs are usually around the Manhattan area. When we select the centroid of the most populous taxi zone as the city center and fit the pickup/drop-off counts of other zones as a function of their distance from the city center, we find that they roughly follow a power-law function, as shown in Fig.~\ref{taxilr}.

Most of the taxi zones fit closely to the regression line, with few following outliers. Three zones that are far from the center but have a relatively higher number of pickup/drop-off counts than most are the three airports of NYC, i.e., LaGuardia, JFK, and Newark International airports, which are not surprising since NYC is well-known as a busy transportation hub. Two zones that have a relatively lower number of pickup/drop-off counts than most at its distance are Rikers Island and Great Kills Park, which are also expected given their lesser popularity. 

This spatial distribution of taxi pickups/drop-offs, i.e., about following a power-law function, agrees with the evidence found in the literature, as presented in~\cite{simplespatialscaling}. In addition, it also suggests that a traffic-based deployment strategy of charging roads will likely be suitable for cities such as New York.

\subsection{Charging Road Assignment \& Model Validation}

Based on the traffic and road network datasets, we want to simulate the scenario where the charging roads are actually deployed and then measure the impact of deployment in terms of the proposed metrics. To this end, on the NYC dataset, we first calculate the distance from the city center to every node in the road network dataset. Then, we define the distance from the city center to each road as the minimum distance from the center to every node of that road. Once the distances from the center to the roads are ready, we assign the charging attribute to each road based on two proposed strategies: uniform deployment and traffic-based power-law deployment. For uniform deployment, each road has a fixed probability of being a charging road, regardless of its distance from the city center. On the other hand, for traffic-based power-law deployment, the probability of charging for a road is a function of its distance from the center. In this experiment, the charging probability is chosen so that on average, 20\% of the roads in NYC will be charging roads. In addition, we assume an average driving speed inside NYC of 20km/h and a constant charging model, i.e., the power that an EV receives equal to the product of the dynamic charging system output power and the time the EV spent on the charging road. Note that in practice, the speed of vehicles may vary, and the charging rate depends on various factors such as the current EV battery level, driving speed, ambient temperature, and road elevation profile. However, since our metrics are aggregated over a large number of trips and this work is one of the first attempts to study the impact of dynamic charging deployment verified on real data, we begin with a basic assumption of constant driving speed and charging rate.   

\begin{figure}[h]
\centering
\captionsetup[subfigure]{font=scriptsize,labelfont=normalsize}
\subfloat[4km\label{sfig:4km}]{%
  \includegraphics[width=0.95\linewidth,keepaspectratio]{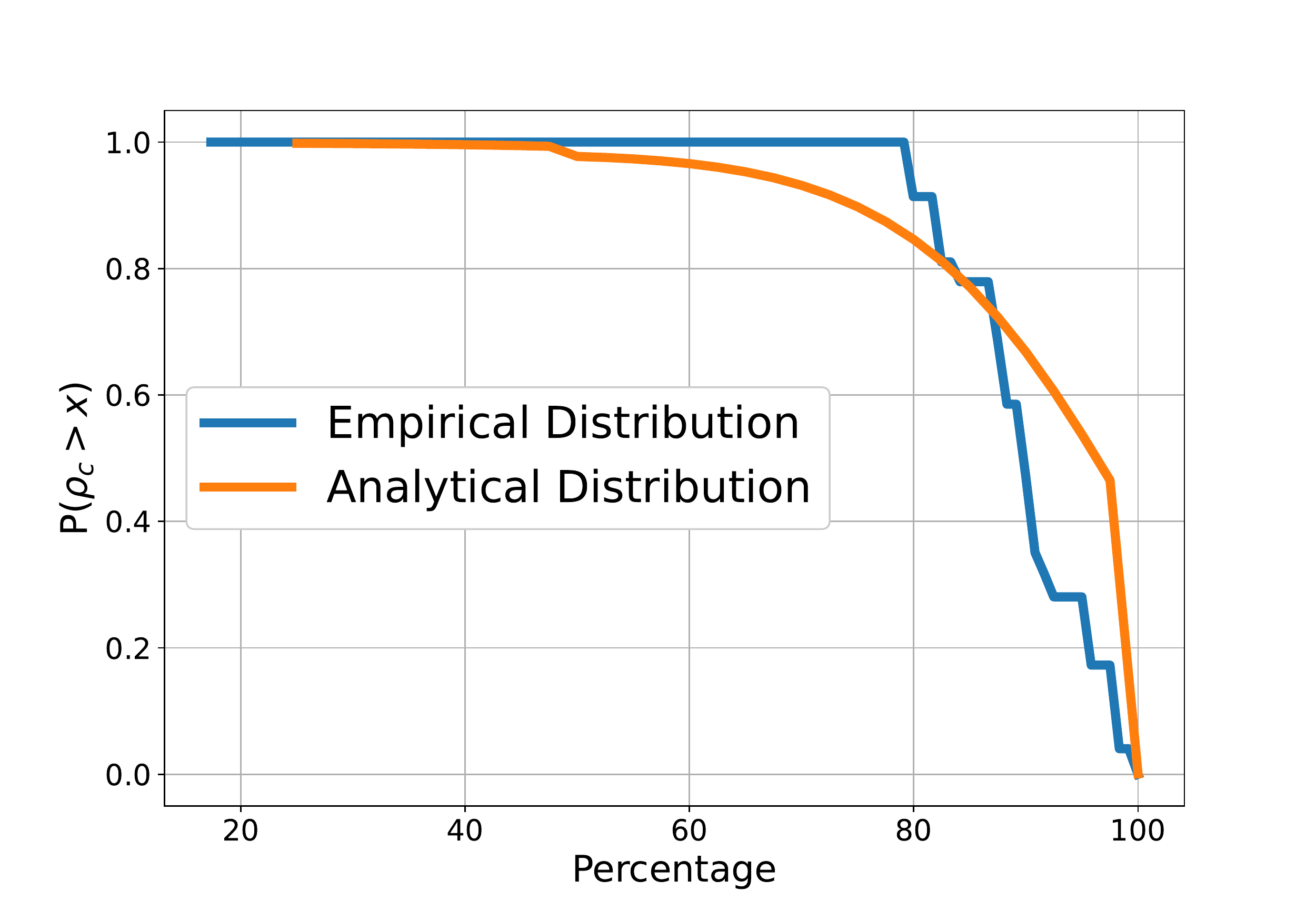}%
}\hfill
\subfloat[7km\label{sfig:7km}]{%
  \includegraphics[width=0.95\linewidth,keepaspectratio]{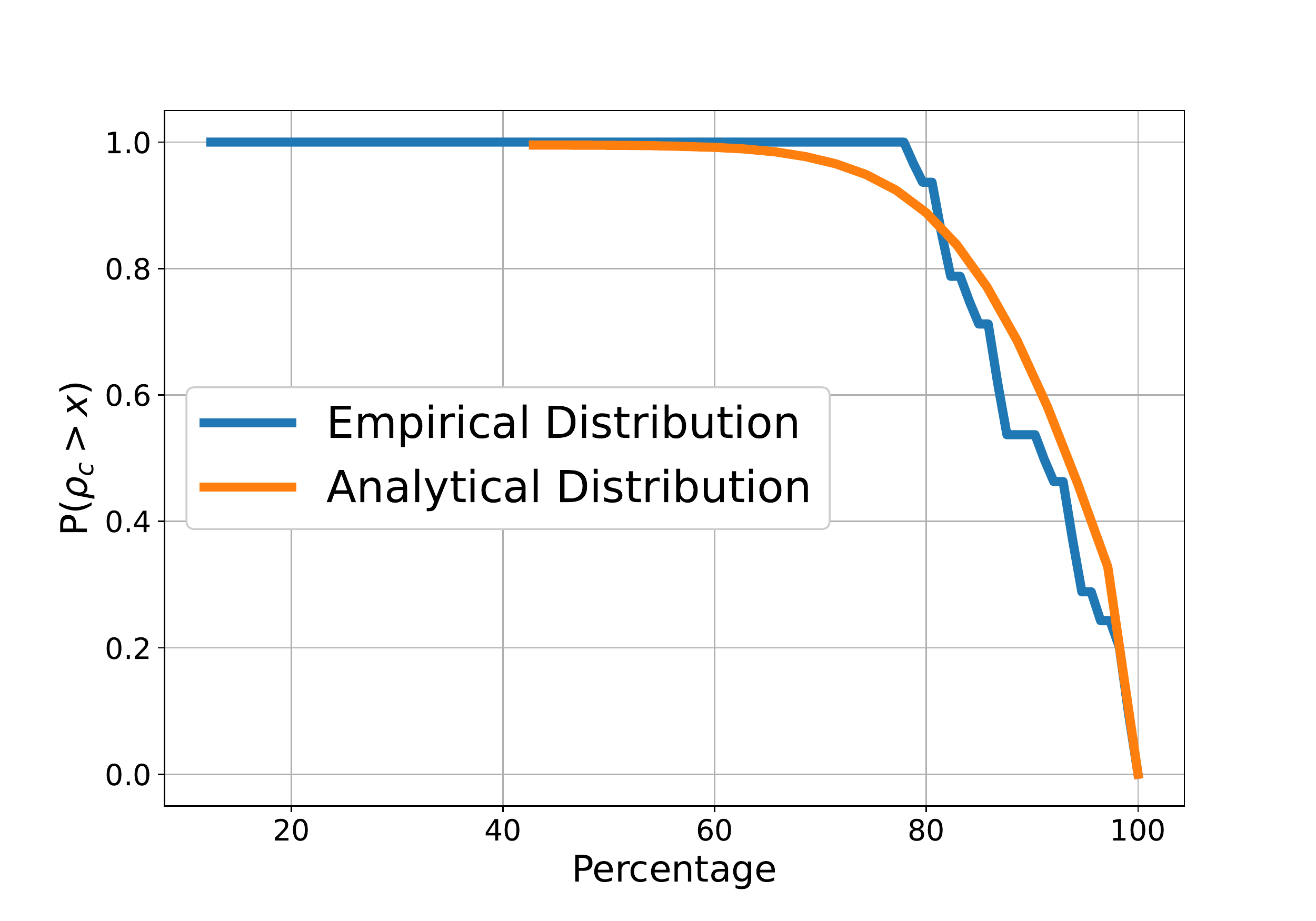}%
}\hfill
\subfloat[10km\label{sfig:10km}]{%
  \includegraphics[width=0.95\linewidth,keepaspectratio]{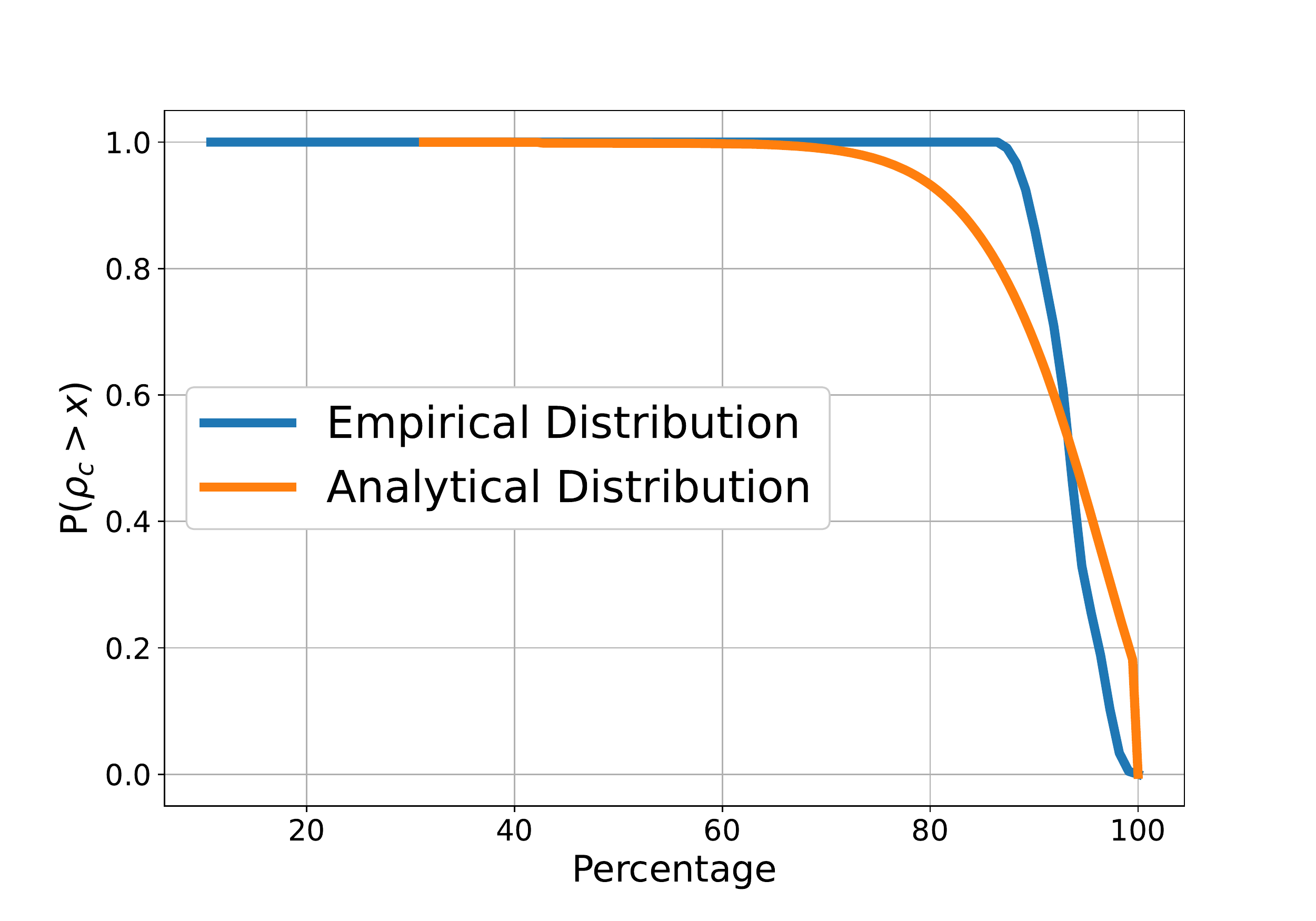}%
}
\caption{The CCDF of road portion traveled on charging roads with different trip lengths across the center of New York City.}
\label{modelvalidation}
\end{figure}

\begin{figure}[h]
\centering
\captionsetup[subfigure]{font=scriptsize,labelfont=normalsize}
\includegraphics[width=0.99\linewidth,keepaspectratio]{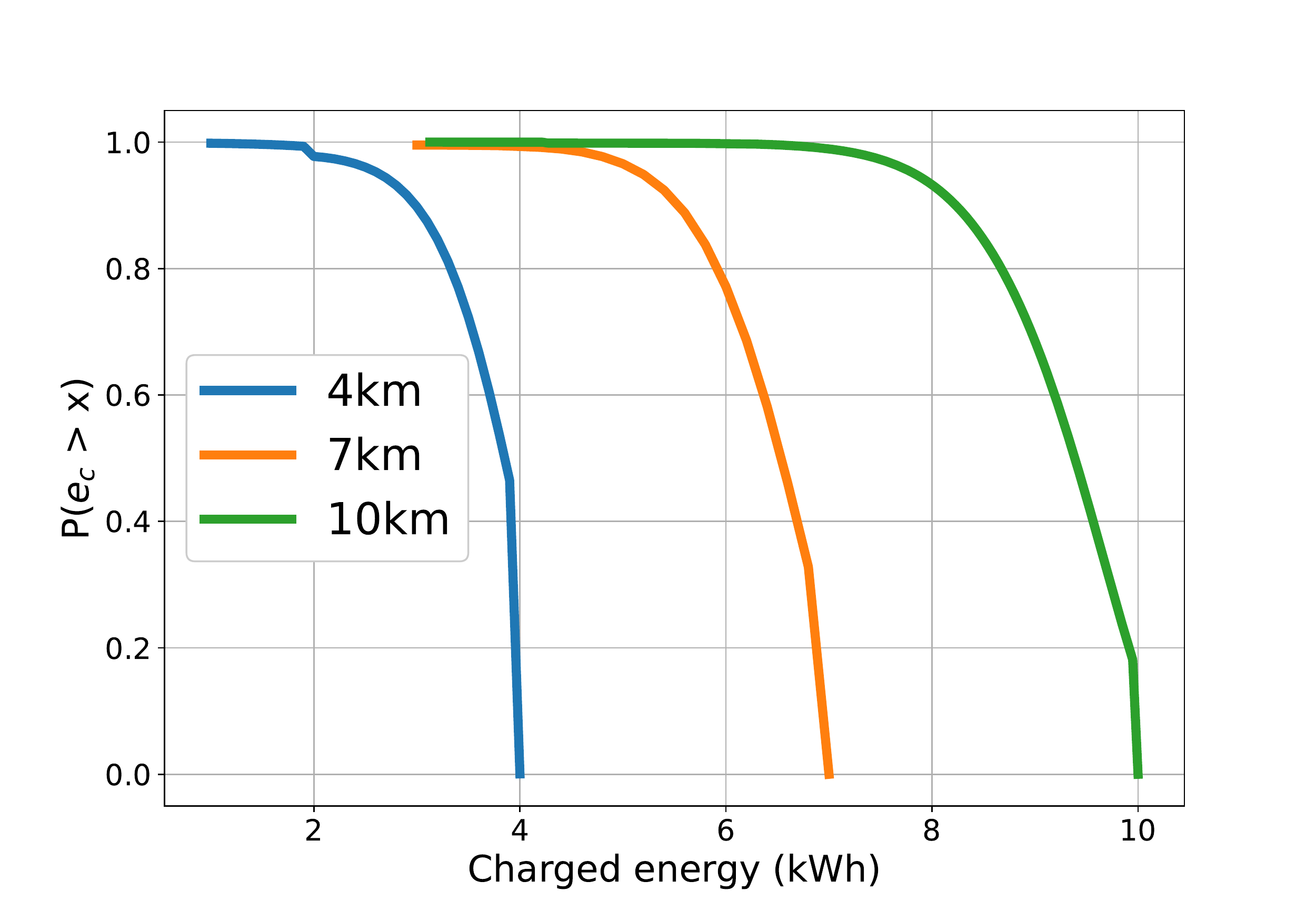}
\caption{The CCDF the energy charged in a trip of different lengths across the center of New York City.}
\label{energychargeddistribution}
\end{figure}

When the charging roads are deployed following a traffic-based strategy, the simulation results for $\rho_c$ with different trip lengths, i.e., 4km, 7km, and 10km across the center of NYC, are shown in Fig.~\ref{modelvalidation}. We see that in all three cases, the empirical distribution that we get from simulating trips on the actual roads of New York closely matches the analytical distribution that we find in section~\ref{deploymentassesment}. In addition, we see from Fig.~\ref{modelvalidation} that in all three cases, $P(\rho_c>80) > 0.8$, indicating that with 20\% of the roads being equipped with dynamic charging as in this case study, trips across the center of NYC will have more than 80\% of the road portion traveling on charging road with a high probability of more than 0.8. Similarly, since $P(\rho_c>40) = 1$ and $P(\rho_c>90) > 0.5$, drivers in NYC will typically commute on more than 40\% of charging road all the time and 90\% of charging road half of the time. This distribution can be used by relevant organizations and policy makers to plan the deployment density of dynamic charging roads. Furthermore, given a trip distance, the distribution of the energy charged in a trip, denoted as $e_c$, can also be derived. In Fig.~\ref{energychargeddistribution}, we show the probability of getting a certain amount of energy through a trip of different lengths. This distribution benefits EV manufacturers in configuring the battery size and suggesting suitable charging schedules to drivers. To gain more insights into the impact of deployment density on urban trips, we simulate traveling scenarios for various EV models in section~\ref{implications}.  

\subsection{Implications to popular EVs models in NYC}
\label{implications}
After developing and validating the model used to compute the distribution of road portions traveled on charging roads, we aim to investigate the impact of deployment density on the travel activities of popular EV models. Specifically, we elaborate on how different deployment schemes, i.e., traffic-based power law deployment and random deployment, at various densities affect the battery levels of common EVs as they commute in NYC. To this end, we select three of the best-selling EV models in the U.S.~\cite{usevsales} whose specifications are given in Table~\ref{EVmodels}.

\begin{table}[h]
\centering
\captionsetup{font=normalsize}
\caption{Popular EV models and their specifications}
\label{EVmodels}
\begin{tabular}{|p{2.5cm}|p{2.5cm}|p{1.5cm}|}
\hline
Model                    & EPA Energy Consumption Rate (kWh/km) & Battery Capacity (kWh) \\ \hline
Tesla Model 3 Range Plus & 0.149129              & 50            \\ \hline
Chevrolet Bolt           & 0.180197              & 60           \\ \hline
Nissan Leaf              & 0.186411              & 40            \\ \hline
\end{tabular}
\end{table}

\begin{figure}[h]
\centering
\captionsetup[subfigure]{font=footnotesize}
\includegraphics[width=0.99\linewidth,keepaspectratio]{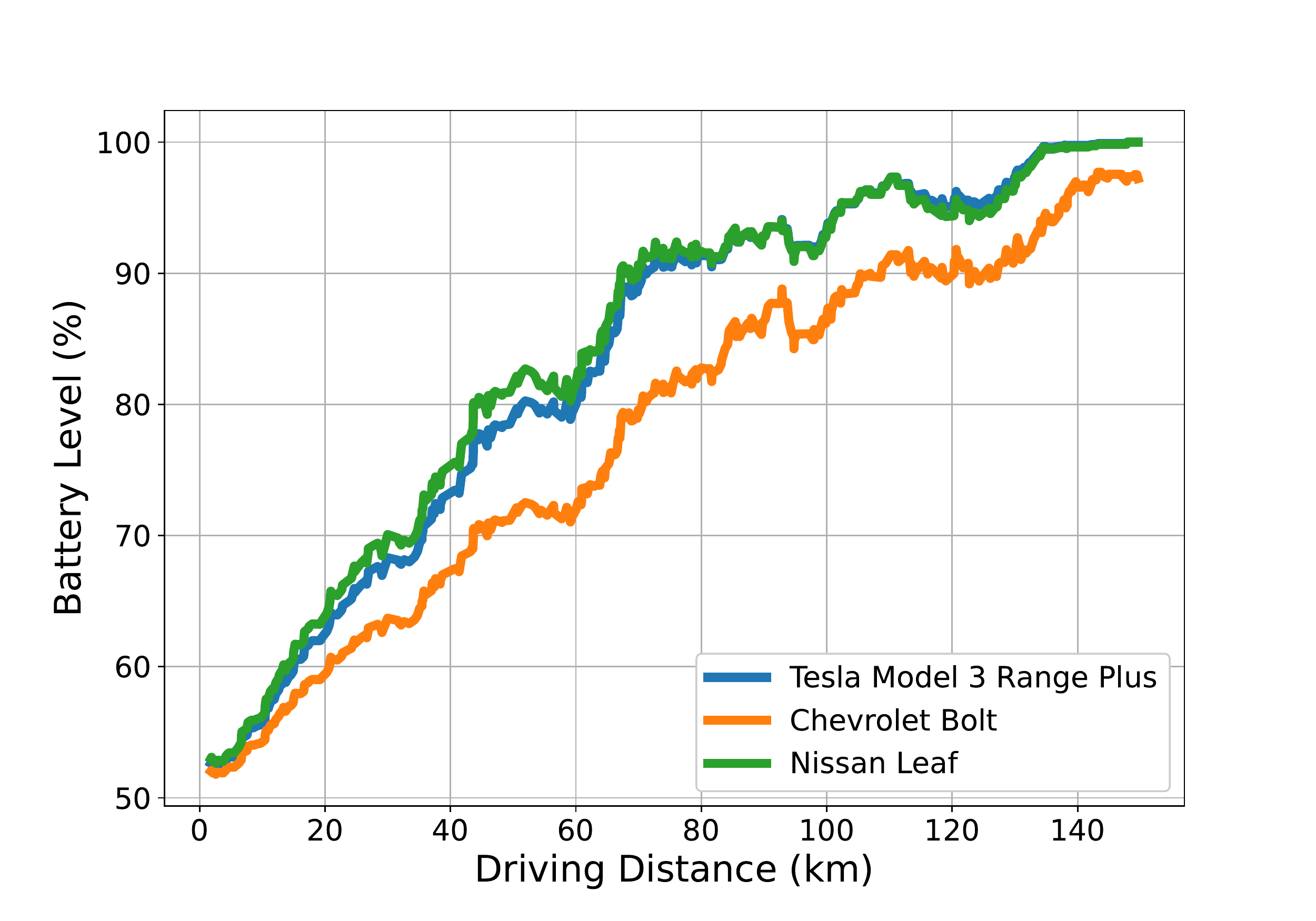}
\caption{Popular EV models traveling in NYC with 20\% charging roads.}
\label{nissanteslachev}
\end{figure}

For all EVs, we assume a starting battery level of 50\% and adopt a simple energy consumption model, i.e., the energy consumption of a trip equals the energy consumption rate multiplied by the length of that trip. Next, based on a survey of popular dynamic charging systems~\cite{reviewofstaticanddynamic}, we assume the power of a typical dynamic charging system to be 20 kW. Then, the energy gain after a trip will be calculated as a product of the dynamic charging system power, the travel time, and the charging portion. Using the actual trip records from the NYC dataset, we estimate the battery levels of the three EVs as they travel in NYC with 20\% of the roads being charging, as shown in Fig.~\ref{nissanteslachev}.

\begin{figure}[h]
\centering
\captionsetup[subfigure]{font=footnotesize}
\includegraphics[width=0.99\linewidth,keepaspectratio]{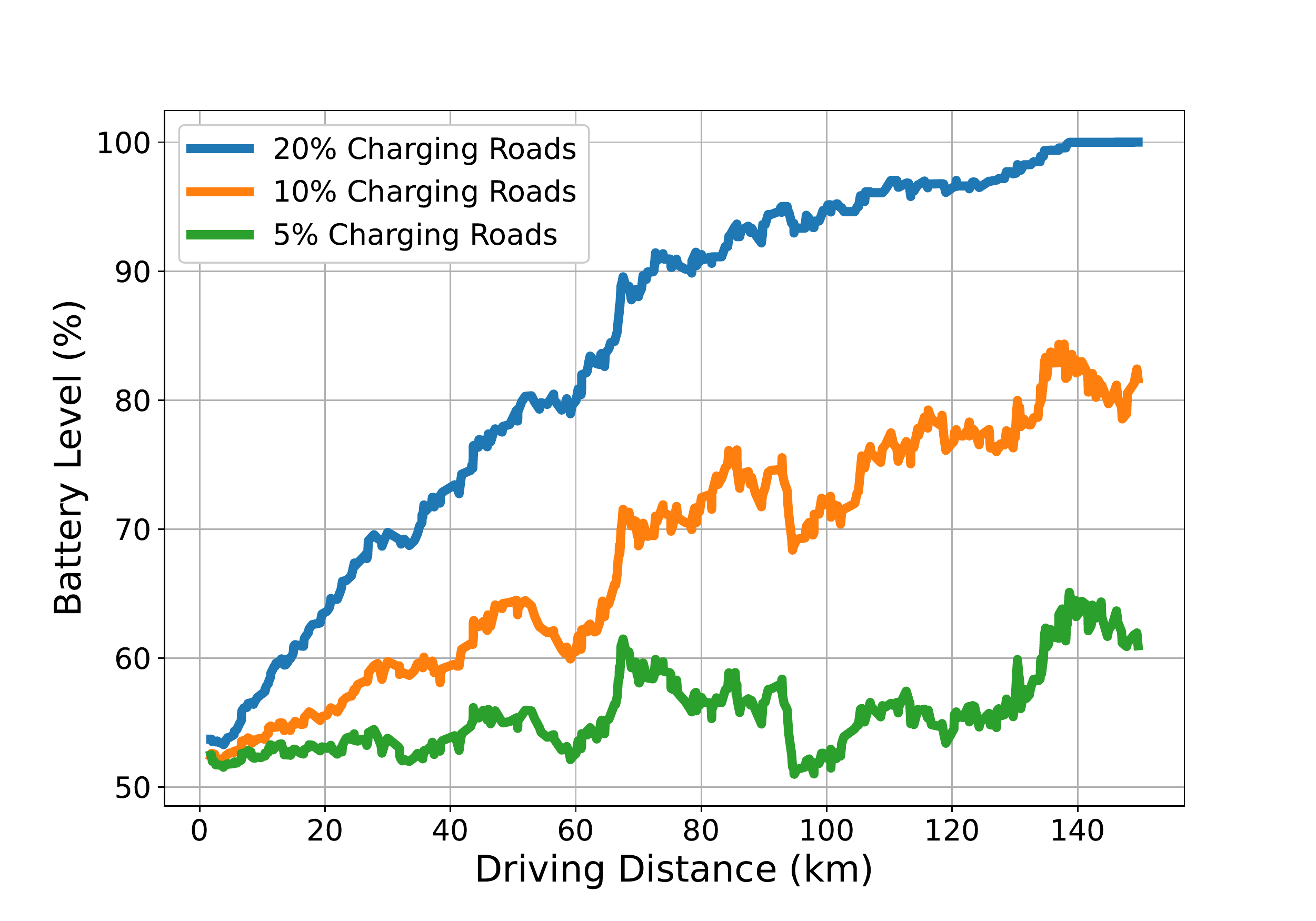}
\caption{Nissan Leaf traveling with power law deployment of charging roads.}
\label{nissanpow20105}
\end{figure}

From Fig.~\ref{nissanteslachev}, we see that all EV models gain approximately 5\%-7\% of their battery levels after every 10 km driving when traveling in NYC with 20\% of the roads being charging. While the Nissan Leaf and the Tesla Model 3 Range Plus have similar battery levels, the Chevrolet Bolt takes a little more time to get fully charged, partly because the Bolt has the biggest battery capacity out of three. Nevertheless, in general, the remarkable gain in battery levels across all 3 EV models suggests a strong capability of dynamic charging systems to power EVs in their everyday commute. Indeed, in Fig.~\ref{nissanpow20105}, the simulated battery level of the Nissan Leaf when traveling with different percentages of charging roads is shown. It is clear that even with only 5\% of charging roads, the Nissan Leaf maintains its battery level throughout its trips. This indicates that EV owners need to visit charging stations much less often with dynamic charging, reducing the demand for building more charging stations. In addition, for transportation and logistic companies, it means their electric fleets can operate continuously with less idle time. 

\begin{figure}[h]
\centering
\captionsetup[subfigure]{font=footnotesize}
\includegraphics[width=0.99\linewidth,keepaspectratio]{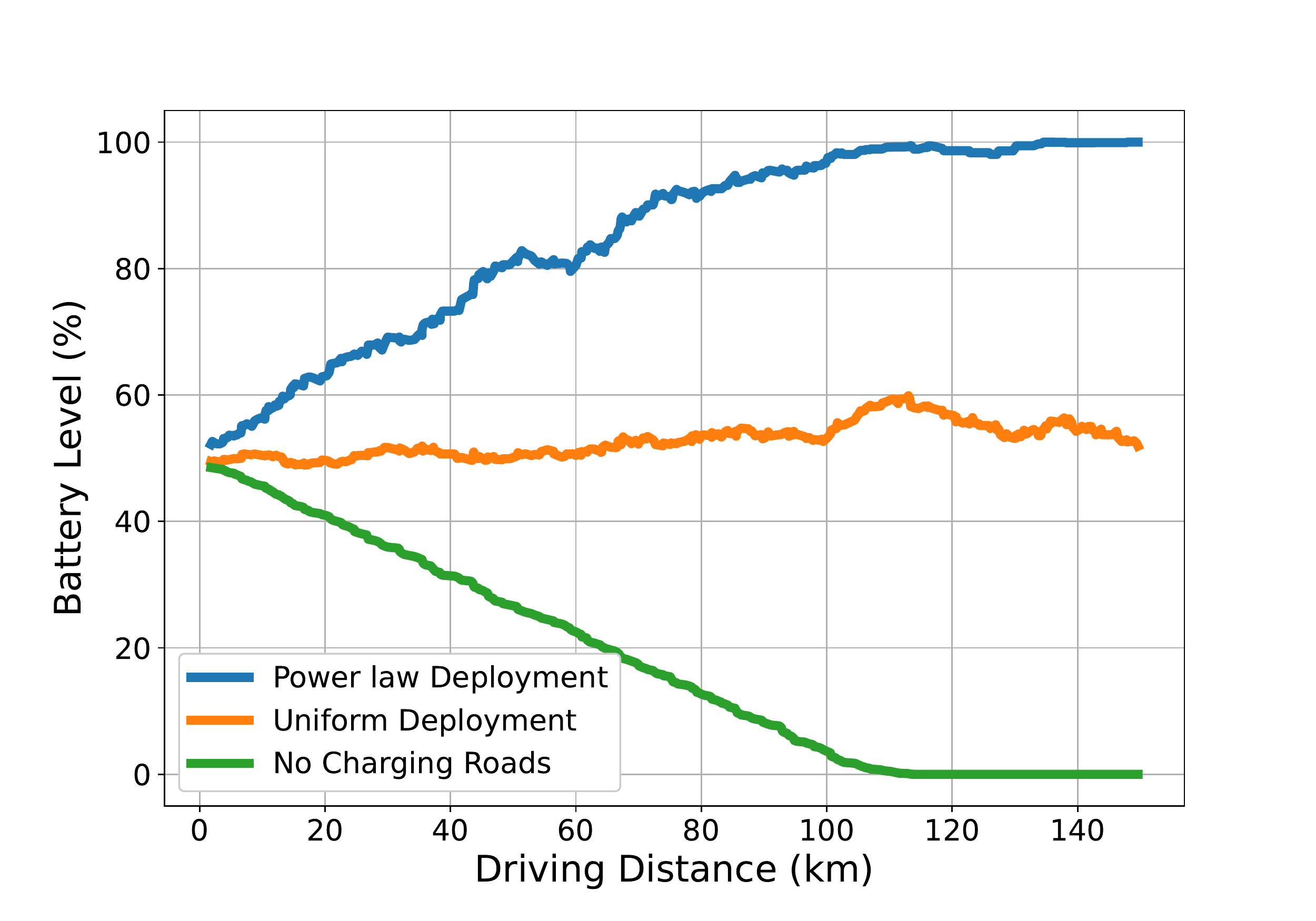}
\caption{Nissan Leaf traveling in NYC with 20\% charging roads.}
\label{nissanpowrand}
\end{figure}

After demonstrating how dynamic charging systems power EVs in their daily trips, we hereby compare the traffic-based power law deployment and the random deployment plans, as presented in Fig.~\ref{nissanpowrand}. The battery level simulation with the Nissan Leaf implies that with the same percentage of charging roads at 20\%, the traffic-based power law deployment scheme way outperforms the random deployment plan in terms of the energy it provides to EVs. For example, from Fig.~\ref{nissanpowrand}, we see that after 100km of driving, while the random deployment only roughly maintains the battery level of the Nissan Leaf, the power law deployment charges the Leaf to its nearly full capacity. Moreover, we observe that after about 110km, the battery of the Leaf will otherwise deplete if there are no charging roads, which underscores the ability of dynamic charging systems to extend the driving range of EVs significantly. 

\subsection{Locally Optimized Deployment Plan for a Dense Neighborhood in Xi'an}
\label{xian}

\begin{figure}[h]
\centering
\captionsetup[subfigure]{font=footnotesize}
\includegraphics[width=0.99\linewidth,keepaspectratio]{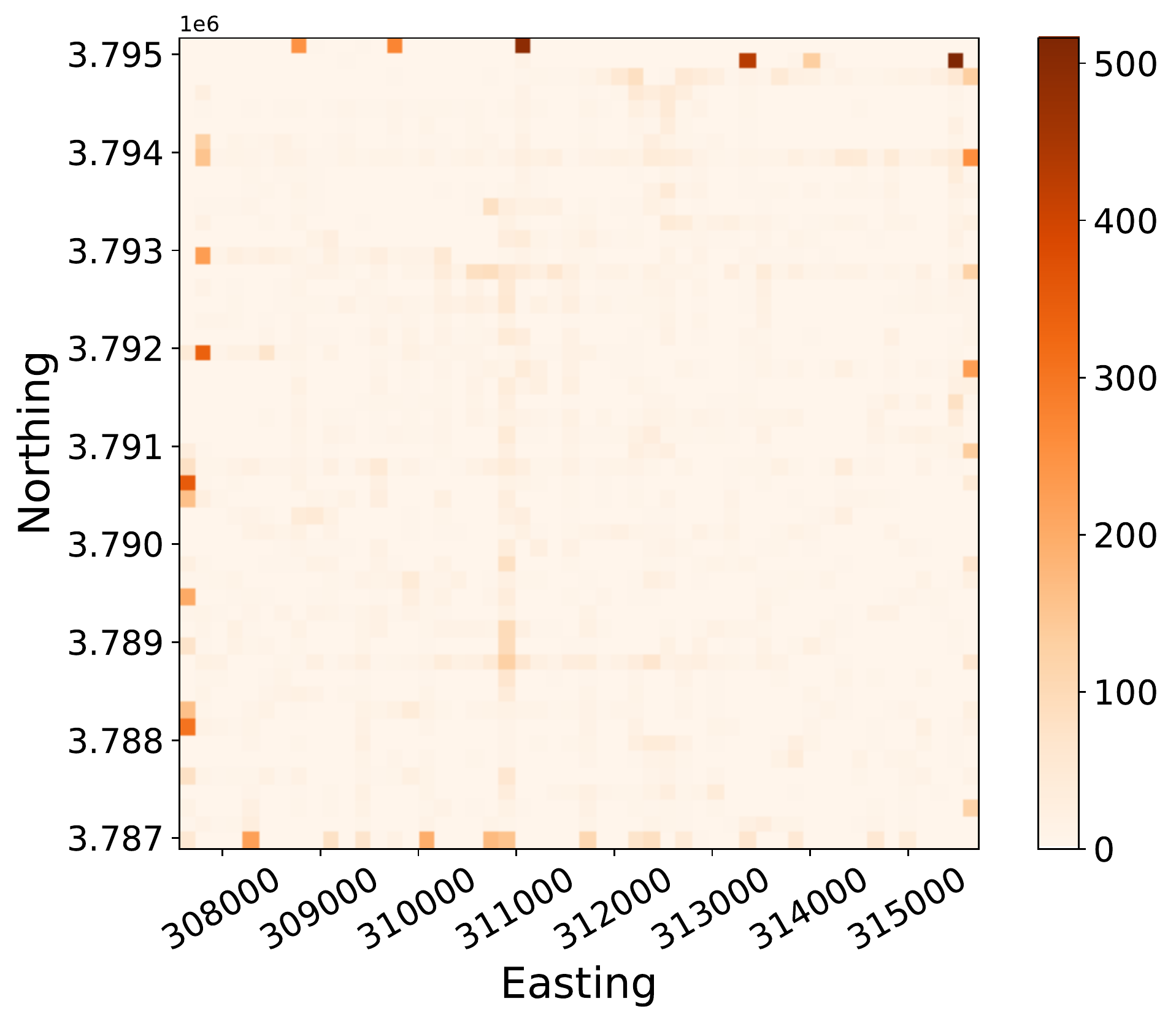}
\caption{Heatmap of pickup/drop-off locations in Xi'an in UTM coordinate system, zone 49.}
\label{heatmapxian}
\end{figure}

In this subsection, we examine the Xi'an dataset, which provides trip records in a dense area of Xi'an. Unlike the previous subsection in which we investigate deployment plans across NYC, in this subsection, we zoom into a populated area of Xi'an to see if we can further optimize the deployment of charging roads inside this dense neighborhood. Similarly, we compare two strategies: the traffic-based deployment plan and the random deployment plan. To explore traffic, we first plot the pickup/drop-off points of the trips on a heatmap, as in Fig~\ref{heatmapxian}.

\begin{figure}[h]
\centering
\captionsetup[subfigure]{font=footnotesize}
\includegraphics[width=0.99\linewidth,keepaspectratio]{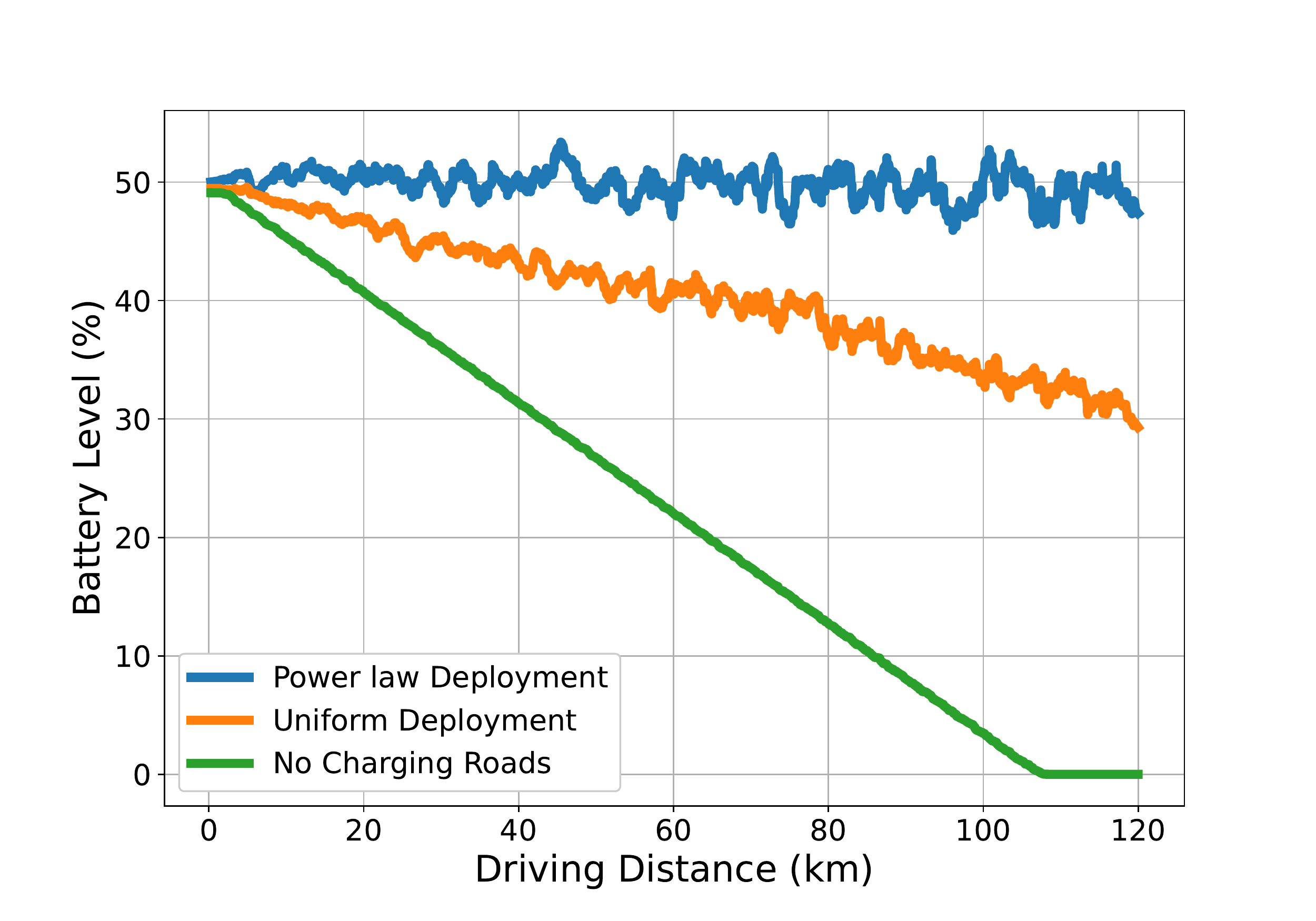}
\caption{Nissan Leaf traveling in Xi'an with 12.74\% charging roads.}
\label{nissanpowrandxian}
\end{figure}

From Fig.~\ref{heatmapxian}, one can see that there are some possible hotspots of pickup/drop-off locations near the edge of the data boundary. For a traffic-based deployment plan in this area, since there is not a clear center and the decreasing trend of traffic from a single center as in the case of NYC cannot be observed, we propose a strategy in which we deploy the charging roads at a decreased density from the top populated traffic clusters. Specifically, we first select the top five most populous clusters of pickup/drop-off locations using a popular density-based clustering algorithm, i.e., DBSCAN~\cite{schubert2017dbscan}. Then, we consider each of those clusters as a mini city center. Next, each road will be assigned a charging probability following a power law function in terms of its minimum distance to the five centers. The idea behind this approach is aligned with the one used in NYC, i.e., to deploy the charging roads where the traffic is dense and then decrease their densities following a power law function. The battery simulation results of the Nissan Leaf traveling in Xi'an with 12.74\% of the roads being charging are presented in Fig.~\ref{nissanpowrandxian}. It is clear that while the traffic-based power law deployment plan roughly retains the Leaf battery level throughout the trips, with the random deployment plan, the Leaf loses nearly 20\% of its battery after 100km of driving.

\subsection{A Comparison with the Gaussian Deployment Plan}

\begin{figure}[h]
\centering
\captionsetup[subfigure]{font=scriptsize,labelfont=normalsize}
\subfloat[5\%\label{sfig:5percent}]{%
  \includegraphics[width=0.98\linewidth,keepaspectratio]{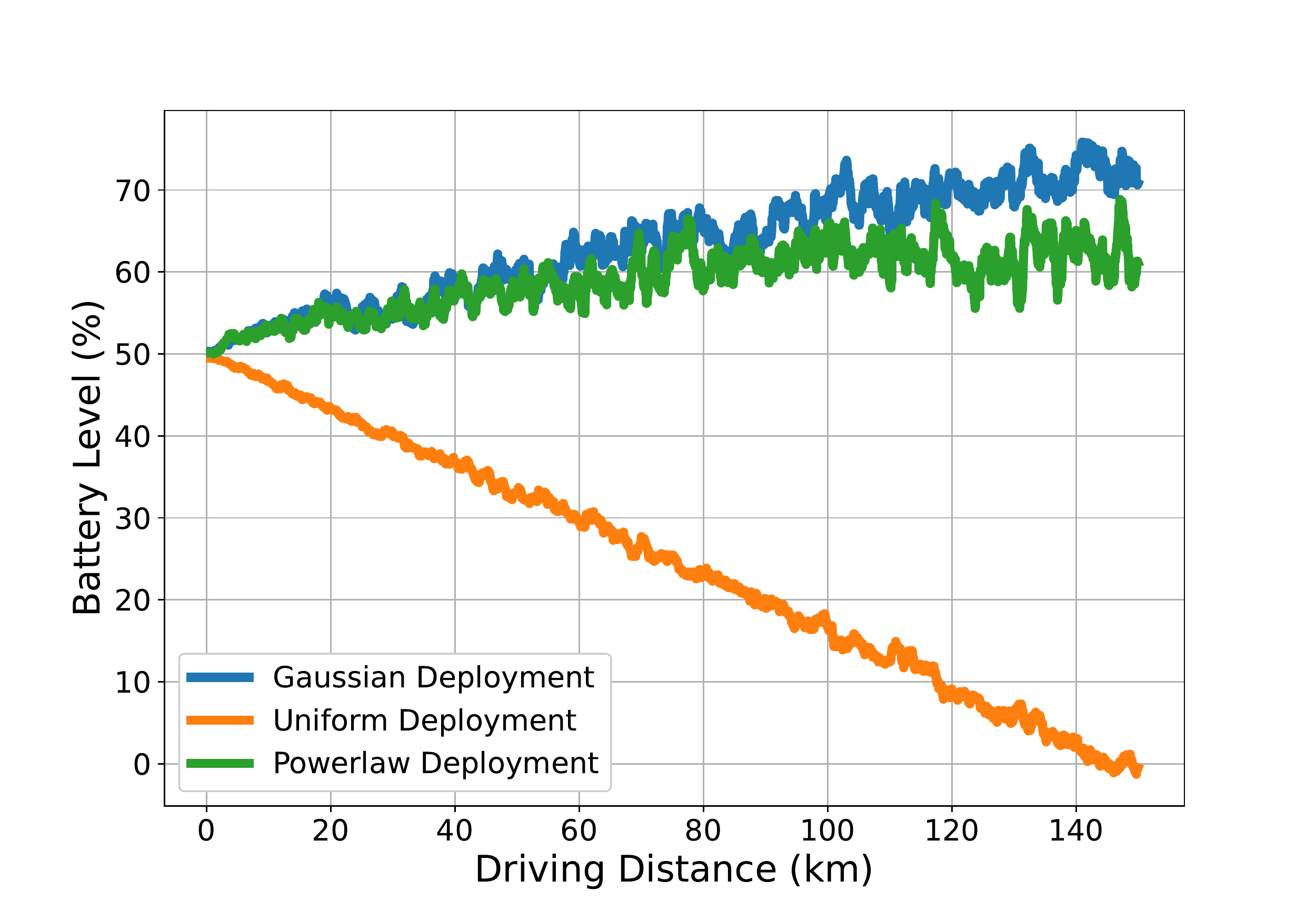}%
}\hfill
\subfloat[15\%\label{sfig:15percent}]{%
  \includegraphics[width=0.98\linewidth,keepaspectratio]{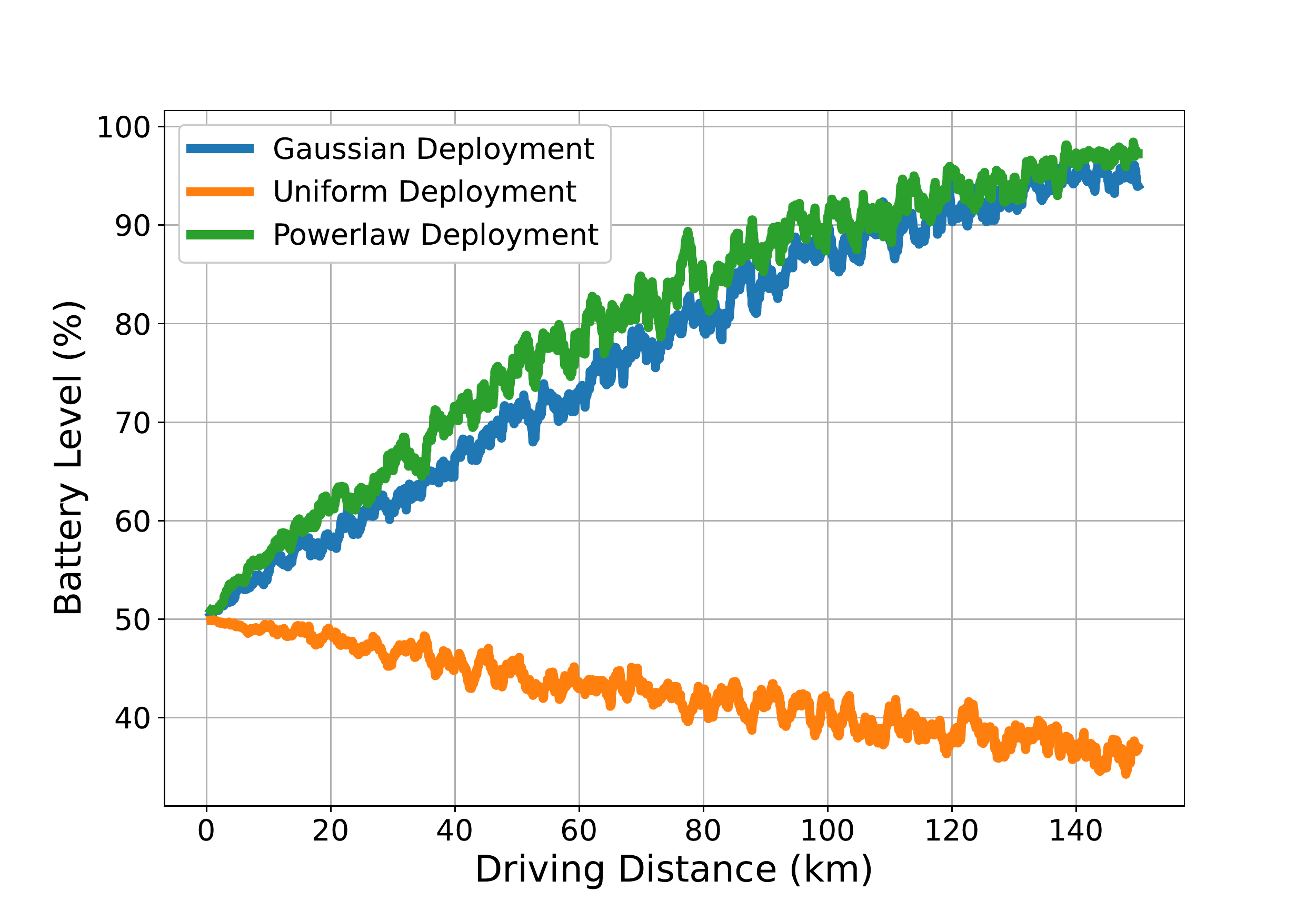}%
}\hfill
\subfloat[25\%\label{sfig:25percent}]{%
  \includegraphics[width=0.98\linewidth,keepaspectratio]{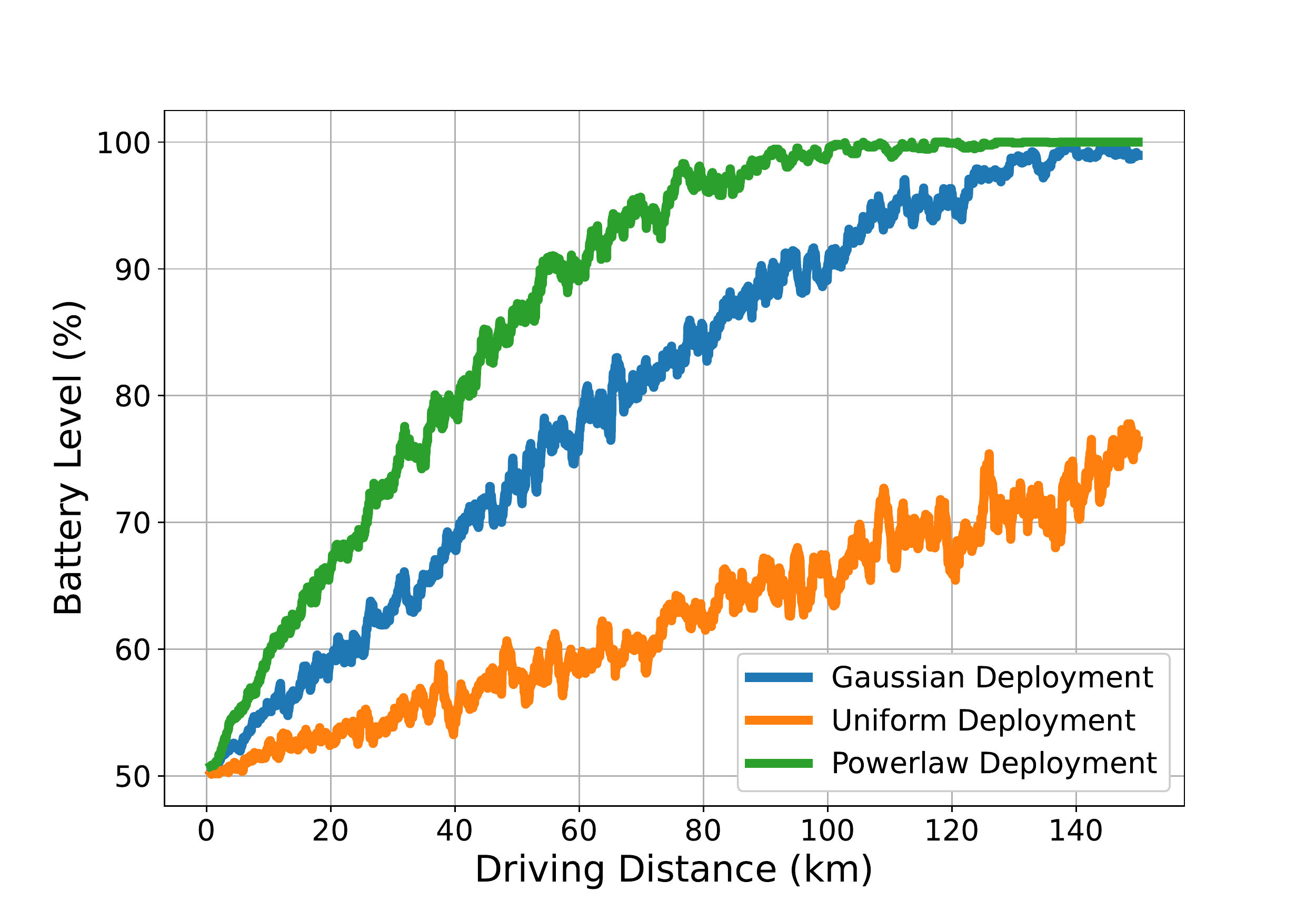}%
}
\caption{Nissan Leaf travelling in NYC with various percentages of charging roads.}
\label{powgau}
\end{figure}

\begin{figure}[ht]
\centering
\captionsetup[subfigure]{font=footnotesize}
\includegraphics[width=0.99\linewidth,keepaspectratio]{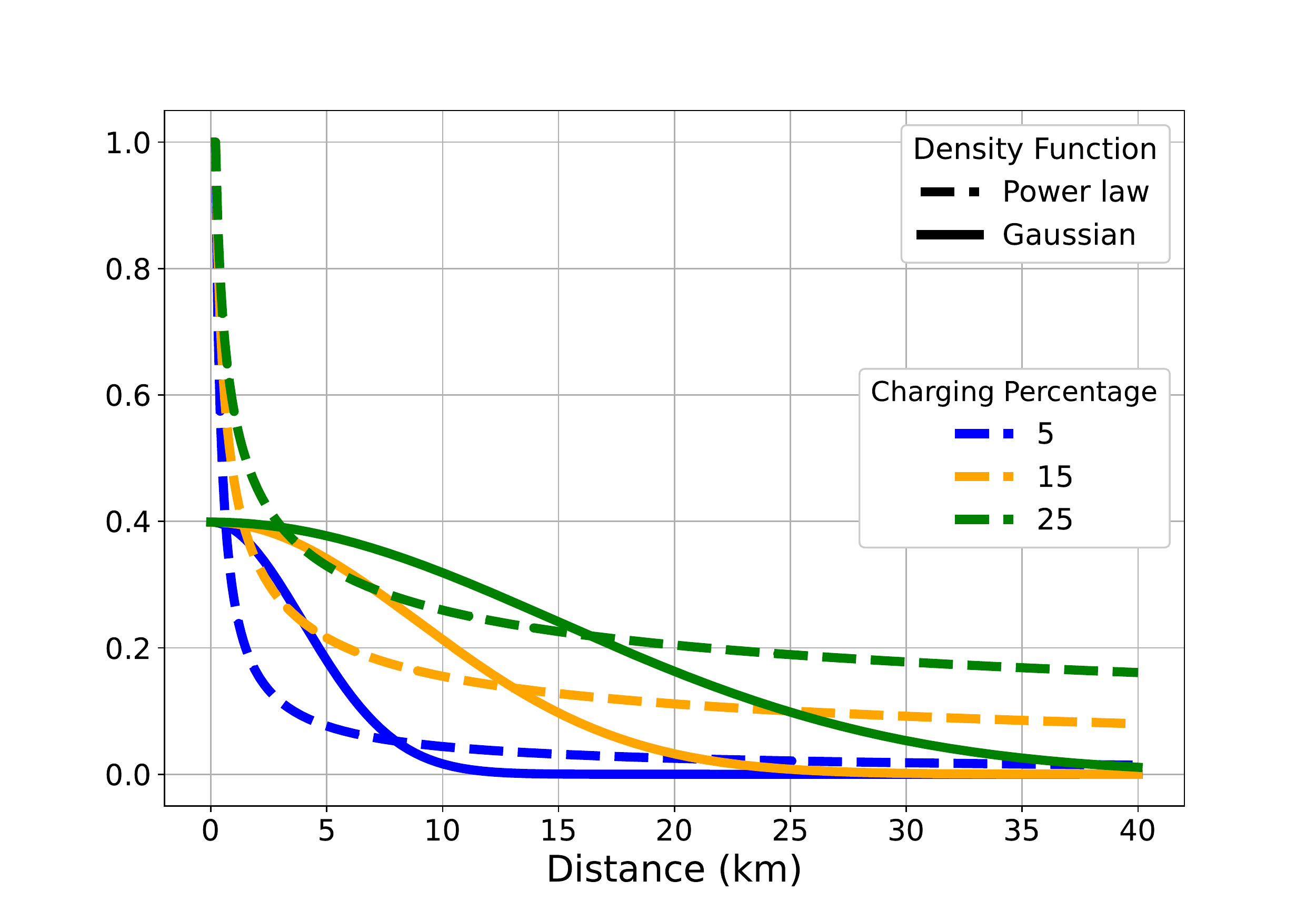}
\caption{Density function according to various road average charging percentage.}
\label{powgaudensity}
\end{figure}

In section~\ref{SpatialDistributionofUrbanTrips} and section~\ref{implications}, we show the evidence that in NYC, where the pickup/drop-off locations follow a power law distribution, a power law deployment plan of charging roads clearly outperforms the uniform one. In this subsection, we further put the power law deployment plan to the test against a much similar strategy, which uses a Gaussian density function, to see if mimicking the distribution of traffic is indeed a good approach. To guarantee a fair comparison, the parameters of both plans are matched to the same average road charging probability. The comparison results of the two plans at different densities are presented in Fig.~\ref{powgau}. One can see that although at low charging density, the Gaussian plan is slightly better than the power law one, when the average road charging probability is above 15\%, the power law strategy apparently outperforms the Gaussian plan in terms of the power each plan provides to EVs. This is because at low charging density, both power law and Gaussian function share a ``long tail", i.e., a zone in which the charging probability of the roads is approximately zero, as shown in Fig.~\ref{powgaudensity}. The only difference is in the area close to the city center, in which the Gaussian function has a higher density, resulting in a slightly better performance. However, when the average road charging probability is above 15\%, the difference between the two functions becomes more substantial. In this case, the power law plan is significantly better since it is closer to the actual traffic. Hence, the comparison suggests that, in general, a traffic-based deployment plan, e.g., the power law plan in the case of NYC, is the best plan to deploy dynamic charging systems in metropolitan cities. 

\section{Concluding Remarks}
In this research, we examined several strategies to deploy dynamic charging systems in metropolitan cities to alleviate the shortcomings of charging stations and thus facilitate vehicle electrification. To compare the strategies, we presented two statistical metrics that capture deployment's impact on urban commuting. Next, we applied these metrics to deduce insights on the changes of battery levels of popular EV models under different deployment scenarios. We found that a traffic-based deployment strategy is not only superior to other deployment schemes in terms of the power provided to EVs, but it is also efficient, e.g., only 5\% of charging roads in NYC can retain the battery levels of EVs without stopping to recharge. This research aims to assist decision makers in public or private organizations in planning the deployment of dynamic charging in urban cities. In the future, when dynamic
charging roads become so popular in metropolitan cities that their costs substantially decrease, this work can also serve as a reference for the deployment in suburban and rural areas. In addition, several directions of future work can be extended from our findings. For example, the system model can be extended to the general PLP so that the results can be applied to cities with non grid-like street networks. In addition, various factors such as the driving speed, traffic flow, congestion, and road elevation profile can be considered to update the metrics and capture the impact of dynamic charging deployment even more precisely. 


\appendices
\section{Proof of Proposition~\ref{propDnvc}}
\label{proofdvc}
\begin{align}
    &\mathbb{P}(D_\mathrm{N-VC} < x) = \mathbb{E}_{S,D}[\mathbb{P}(D_\mathrm{N-VC} < x)| S, D)] 
  \nonumber\\&= \mathbb{E}_{S,D}[1 - \mathbb{P}(D_\mathrm{N-VC} > x)| S, D)]
    \nonumber\\& = \mathbb{E}_{S,D}[1 - \mathbb{P}(D_\mathrm{N-VC} > x)| S > D)\mathbbm{1}\{S > D\} \nonumber\\&- 
     \mathbb{P}(D_\mathrm{N-VC} > x)| S < D)\mathbbm{1}\{S < D\}] 
\end{align}

\section{Closed form of Proposition~\ref{propDnvcSD}}
\label{grclosedform}
Assume that $S<D$, we now derive the closed form for $\int_{s}^{s+x} g(r) {\rm d}r$.
\begin{align}
    &\int_{s}^{s+x} g(r) {\rm d}r \nonumber\\&= \int_{s}^{s+x} \biggl(\frac{\mid r \mid}{r_{\rm min}}\biggr)^{-\alpha}\mathbbm{1}\{\mid r \mid> r_{\rm min}\} + \mathbbm{1}\{\mid r \mid< r_{\rm min}\} {\rm d}r
    \nonumber\\&=\int_{s}^{s+x} \biggl(\frac{\mid r \mid}{r_{\rm min}}\biggr)^{-\alpha}(\mathbbm{1}\{r > r_{\rm min}\} + \mathbbm{1}\{r < -r_{\rm min}\}) \nonumber\\&+ \mathbbm{1}\{-r_{\rm min}< r < r_{\rm min}\} {\rm d}r
    \nonumber\\&=\int_{s}^{{\rm min}(s+x,-r_{\rm min})} \biggl(\frac{- r }{r_{\rm min}}\biggr)^{-\alpha} \mathbbm{1}\{s < -r_{\rm min}\} {\rm d}r 
    \nonumber\\&+\int_{{\rm max}(-r_{\rm min},s)}^{{\rm min}(s+x,r_{\rm min})} \mathbbm{1}\{-r_{\rm min}-x <s < r_{\rm min}\} {\rm d}r
    \nonumber\\&+\int_{{\rm max}(r_{\rm min},s)}^{s+x} \biggl(\frac{ r }{r_{\rm min}}\biggr)^{-\alpha} \mathbbm{1}\{s +x > r_{\rm min}\} {\rm d}r
    \nonumber\\&=\frac{-r_{\rm min}}{1-\alpha}\biggl(\biggl(\frac{-{\rm min}(s+x,-r_{\rm min})}{r_{\rm min}}\biggr)^{1-\alpha}-\biggl(\frac{-s}{r_{\rm min}}\biggr)^{1-\alpha}\biggr)\nonumber\\&\times\mathbbm{1}\{s < -r_{\rm min}\}
    + ({\rm min}(s+x,r_{\rm min}) - {\rm max}(-r_{\rm min},s))\nonumber\\&\times\mathbbm{1}\{-r_{\rm min}-x <s < r_{\rm min}\}
    \nonumber\\&+ \frac{r_{\rm min}}{1-\alpha}\biggl(\biggl(\frac{s+x}{r_{\rm min}}\biggr)^{1-\alpha}-\biggl(\frac{{\rm max}(s,r_{\rm min})}{r_{\rm min}}\biggr)^{1-\alpha}\biggr)\nonumber\\&\mathbbm{1}\{s +x > r_{\rm min}\}.
\end{align}

\section{Proof of proposition~\ref{propX1}}
\label{proofx1}
\begin{align}
    &\mathbb{P}(X_1 < x|S,D, S<D) \nonumber\\&=
     {\scriptstyle \mathbb{P}(D_\mathrm{N-VC} - D_\mathrm{N-VNC} < x | D_\mathrm{N-VNC} < D_\mathrm{N-VC} < d_v,S<D) }
    \nonumber\\& = \frac{\mathbb{P}(D_\mathrm{N-VC} - x < D_\mathrm{N-VNC} < D_\mathrm{N-VC} < d_v)}{\mathbb{P}(D_\mathrm{N-VNC} < D_\mathrm{N-VC} < d_v)}
     \nonumber\\&= {\textstyle \frac{\mathbb{E}_{D_\mathrm{N-VC}}[\mathbb{P}(t - x < D_\mathrm{N-VNC} < t < d_h)|D_\mathrm{N-VC} = t, S<D]}{\mathbb{E}_{D_\mathrm{N-VC}}[\mathbb{P}(D_\mathrm{N-VNC} < t < d_h)|D_\mathrm{N-VC} = t, S<D]} }
\end{align}

\section{Derivation of $D_n$ and $\rho_c$}
\label{proofDnDc}

After calculating $\mathbb{P}(L_{3,5}|S,D)$, we move on to $\mathbb{P}(D_n < x|L_{3,5},S,D)$ as follows: 
\begin{align*}
     &\mathbb{P}(D_n < x|L_{3,5},S,D) \\&= \mathbb{P}(D_n < x|T_{3,4,1},T_{3,3,1},T_{3,2,3},T_{3,1,4},T_3,S,D) \\&= 
     {\scriptscriptstyle \mathbb{P}(D_\mathrm{N-HC} < x| D_\mathrm{N-HC} < d_v, D_\mathrm{N-HNC} < D_\mathrm{N-HC}, D_\mathrm{N-HNC} + d_h>D_\mathrm{N-HC},S,D) } \\&\times\mathbbm{1}\{x <d_v\} + \mathbbm{1}\{x> d_v \}
     \\&={\scriptstyle \frac{\mathbb{E}_{D_\mathrm{N-HNC}}[\mathbb{P}(D_\mathrm{N-HNC}<D_\mathrm{N-HC}<\mathrm{min}(D_\mathrm{N-HNC}+d_h,x)|D_\mathrm{N-HNC},S,D)]}{\mathbb{E}_{D_\mathrm{N-HNC}}[\mathbb{P}(D_\mathrm{N-HNC}<D_\mathrm{N-HC}<\mathrm{min}(D_\mathrm{N-HNC}+d_h,d_v)|D_\mathrm{N-HNC},S,D)]} }
     \\&\times\mathbbm{1}\{x <d_v\} + \mathbbm{1}\{x> d_v \}
     \\&={\textstyle \frac{\int_{0}^{x} (F_{D_\mathrm{N-HC}|S,D}(\mathrm{min}(t+d_h,x)|s,d,s<d) - F_{D_\mathrm{N-HC}|S,D}(t|s,d,s<d)) }{\int_{0}^{d_v} (F_{D_\mathrm{N-HC}|S,D}(\mathrm{min}(t+d_h,d_v)|s,d,s<d) - F_{D_\mathrm{N-HC}|S,D}(t|s,d,s<d))} } \\&\times {\textstyle \frac{f_{D_\mathrm{N-HNC}|S,D}(t|s,d,s<d) {\rm d}t}{f_{D_\mathrm{N-HNC}|S,D}(t|s,d,s<d) {\rm d}t}}
     \times\mathbbm{1}\{x <d_v\} + 
     \mathbbm{1}\{x> d_v \}.
\end{align*}
Similarly, we can calculate $\mathbb{P}(\rho_c < x|L_{3,5},S,D)$.
\begin{align*}
     &\mathbb{P}(\rho_c < x|L_{3,5},S,D) \\&= \mathbb{P}(\rho_c < x|T_{3,4,1},T_{3,3,1},T_{3,2,3},T_{3,1,4},T_3,S,D) =\\&
     {\scriptscriptstyle \mathbb{P}(d_h + d_v - D_\mathrm{N-HC} < x| D_\mathrm{N-HC} < d_v, D_\mathrm{N-HNC} < D_\mathrm{N-HC}, D_\mathrm{N-HNC} + d_h>D_\mathrm{N-HC},S,D) }
     \\&\times\mathbbm{1}\{d_h<x <d_h+d_v\} + \mathbbm{1}\{x>d_h+ d_v \}
     \\&={\textstyle \frac{\mathbb{P}(\mathrm{max}(d_h+d_v-x,D_\mathrm{N-HNC}) < D_\mathrm{N-HC} < \mathrm{min}(d_v,D_\mathrm{N-HNC}+d_h)|S,D )}{\mathbb{P}(D_\mathrm{N-HNC} < D_\mathrm{N-HC} < \mathrm{min}(d_v,D_\mathrm{N-HNC}+d_h)|S,D)} }\\&\times\mathbbm{1}\{d_h<x <d_h+d_v\}
     + \mathbbm{1}\{x>d_h+ d_v \} =
     \\& {\scriptstyle \frac{\mathbb{E}_{D_\mathrm{N-HNC}}[\mathbb{P}(\mathrm{max}(d_h+d_v-x,D_\mathrm{N-HNC})<D_\mathrm{N-HC}<\mathrm{min}(d_v,D_\mathrm{N-HNC}+d_h)|D_\mathrm{N-HNC},S,D)]}{\mathbb{E}_{D_\mathrm{N-HNC}}[\mathbb{P}(D_\mathrm{N-HNC}<D_\mathrm{N-HC}<\mathrm{min}(d_v,D_\mathrm{N-HNC}+d_h)|D_\mathrm{N-HNC},S,D)]} }
     \\&\times\{d_h<x <d_h+d_v\}+ \mathbbm{1}\{x>d_h+ d_v \} =  
     \\&{\scriptstyle \frac{\int_{\mathrm{max}(d_v-x,0)}^{d_v} (F_{D_\mathrm{N-HC}|S,D}(\mathrm{min}(d_v,t+d_h)|s<d) - F_{D_\mathrm{N-HC}|S,D}(\mathrm{max}(d_h+d_v-x,t)|s<d)) }{\int_{0}^{d_v} (F_{D_\mathrm{N-HC}|S,D}(\mathrm{min}(d_v,t+d_h)|s,d,s<d) - F_{D_\mathrm{N-HC}|S,D}(t|s,d,s<d)) }} \\&\times {\textstyle \frac{f_{D_\mathrm{N-HNC}|S,D}(t|s,d,s<d) {\rm d}t}{f_{D_\mathrm{N-HNC}|S,D}(t|s,d,s<d) {\rm d}t}}\mathbbm{1}\{d_h<x <d_h+d_v\}+ \mathbbm{1}\{x>d_h+ d_v \}.
\end{align*}

\ifCLASSOPTIONcaptionsoff
  \newpage
\fi

\bibliographystyle{IEEEtran}
\bibliography{hokie-HD}

\begin{IEEEbiography} 
[{\includegraphics[width=1in,height=1.25in,clip,keepaspectratio]{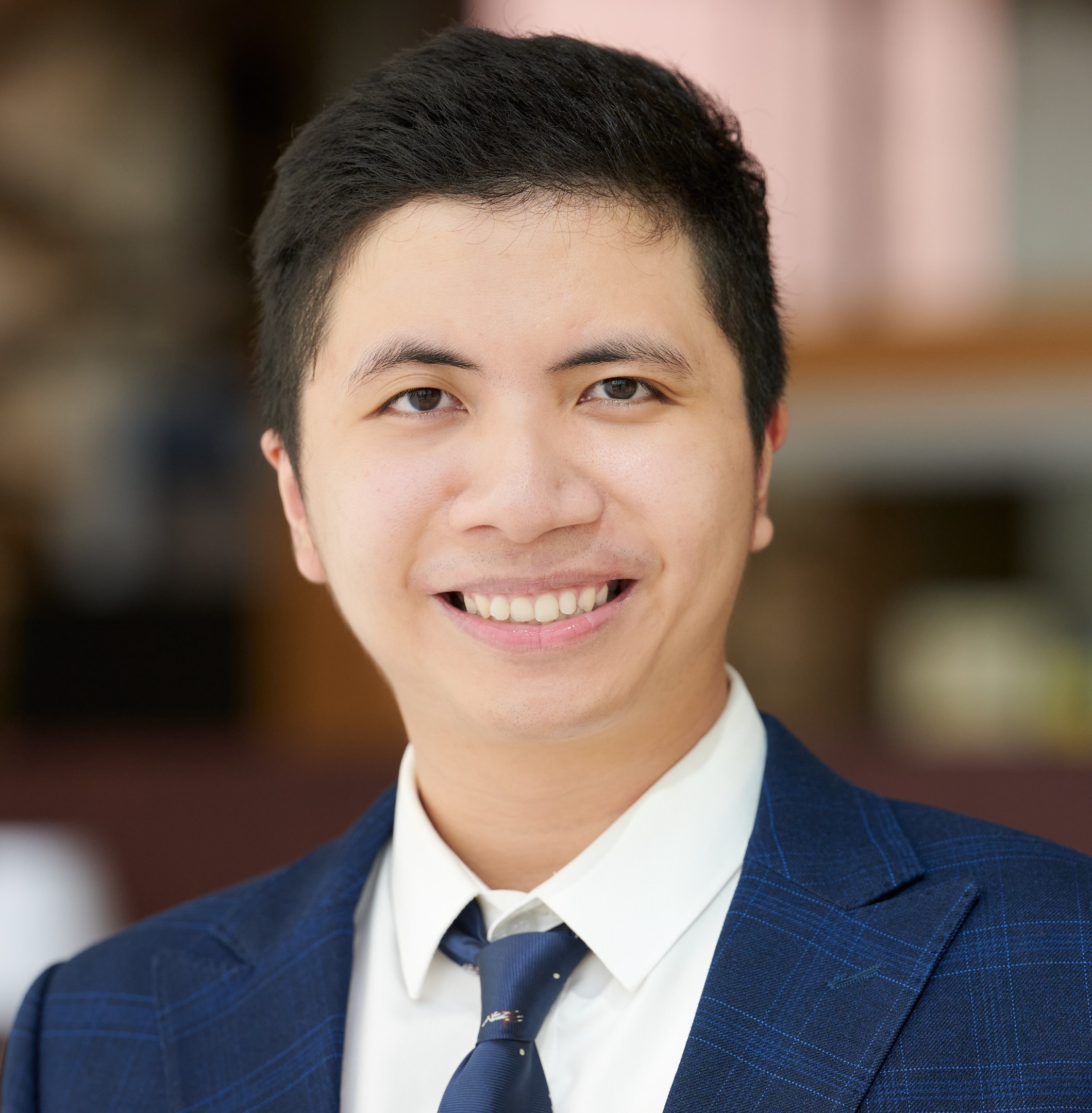}}]{Duc Minh Nguyen}
(Student Member, IEEE) was born in Hanoi, Vietnam. He received the M.S. degree in Electrical and Computer Engineering from King Abdullah University of Science and Technology (KAUST), Saudi Arabia, and the B.Eng. degree in Mobile Systems Engineering from Dankook University, Republic of Korea, in 2020 and 2018, respectively. He is currently pursuing the Ph.D. degree with the Electrical and Computer Engineering Department, King Abdullah University of Science and Technology, Thuwal, Saudi Arabia. His research interests include sustainable transportation and machine learning.
\end{IEEEbiography}

\begin{IEEEbiography}
[{\includegraphics[width=1in,height=1.25in,clip,keepaspectratio]{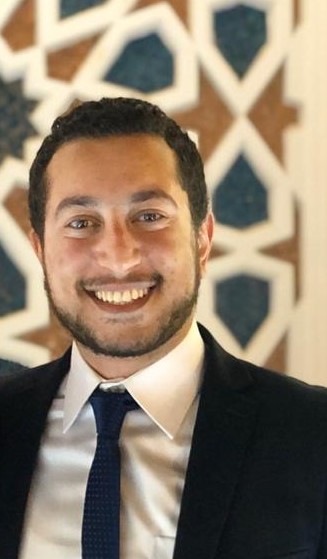}}]{Mustafa A. Kishk}
(Member, IEEE) is an Assistant Professor at Maynooth University, Ireland. From 2019 to 2022 he was a postdoctoral research fellow at KAUST. He received his Ph.D degree from Virginia Tech in 2018, his M.Sc. and B.Sc. degrees from Cairo University in 2015 and 2013, respectively, all in Electrical Engineering. His research interests include UAV communications, satellite communications, and global connectivity for rural and remote areas.
\end{IEEEbiography}

\begin{IEEEbiography}
[{\includegraphics[width=1in,height=1.25in,clip,keepaspectratio]{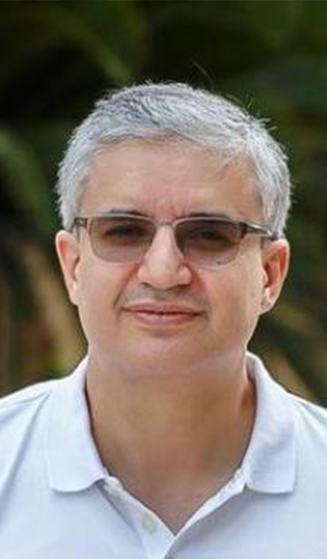}}]
{Mohamed-Slim Alouini}
(Fellow, IEEE) was born in Tunis, Tunisia. He received the Ph.D. degree in Electrical Engineering from the California
Institute of Technology (Caltech), Pasadena, CA, USA, in 1998. He served as a faculty member in the University of Minnesota, Minneapolis, MN, USA, then in the Texas A\&M University at Qatar, Education City, Doha, Qatar before joining King Abdullah University of Science and Technology (KAUST), Thuwal, Makkah Province, Saudi Arabia as a Professor of Electrical Engineering in 2009. His current research interests include the modeling, design, and performance analysis of wireless communication systems.
\end{IEEEbiography}

\end{document}